\documentclass{iopjournal}
\usepackage{amsmath}
\usepackage{orcidlink}
\usepackage{hyperref}
\usepackage{graphicx}
\usepackage{dcolumn}
\usepackage{bm}
\usepackage{wasysym}
\usepackage{hyperref}
\usepackage[mathlines]{lineno}
\usepackage{xcolor}
\usepackage{dsfont}
\usepackage{booktabs}
\usepackage{gensymb}
\usepackage{enumitem}
\usepackage{multirow}
\usepackage{booktabs}
\usepackage{array}
\newcolumntype{P}[1]{>{\centering\arraybackslash}p{#1}}
\usepackage{amssymb}
\usepackage{adjustbox}
\usepackage[normalem]{ulem}

\newcommand{\iu}{\mathrm{i}\mkern1mu}
\newcommand{\du}{\mathrm{d}}

\begin{document}

\title{Boson Stars in $D \ge 4$ Dimensions: Stability, Oscillation Frequencies, and Dynamical Evolutions}

\author{Gareth Arturo Marks$^{1*}$ \orcidlink{0009-0003-3160-9337}, Abdullah Al Zaif$^{1,2}$\orcid{0009-0009-6350-6854}  }

\affil{$^*$Author to whom any correspondence should be addressed.}

\affil{$^1$DAMTP, Centre for Mathematical Sciences,
University of Cambridge, Wilberforce Road, Cambridge CB3 0WA, UK}
\affil{$^2$Sidney Sussex College, Cambridge CB2 3HU, UK}

\email{gam54@cam.ac.uk}


\begin{abstract}
We construct spherically symmetric boson star solutions in $D \in \{4,5,6\}$ spacetime dimensions, considering the effects of both a quartic self-interaction term and a solitonic potential. We then perform a perturbative analysis, generalizing the pulsation equations to arbitrary dimension and potential and hence demonstrating the existence of radially stable higher-dimensional boson star solutions. {In particular, we find stable solutions for $D \in \{5,6\}$ with a quartic self-interaction term and for $D = 5$ with a solitonic potential.} We supplement these linear results with perturbed and unperturbed nonlinear dynamical evolutions in spherical symmetry, obtained using a dimensional reduction that allows us to evolve spacetimes with any number of background dimensions using the same numerical framework, while preserving the full gauge freedom of standard approaches to numerical relativity. The results of these evolutions indicate that the solutions we identify as perturbatively stable are indeed generally stable to nonlinear spherical dynamics. 
\end{abstract}

\section{Introduction}
Boson stars (BSs) are self-gravitating configurations composed of a massive complex field, providing simple exotic-compact-object models that are numerically tractable to dynamical evolution \cite{Liebling_2023}. First introduced by Kaup \cite{Kaup:1968} as well as Ruffini and Bonazzola \cite{Ruffini:1969qy}, in recent years BSs have received attention in a variety of capacities. This includes study as potential sources of gravitational waves, through mergers involving two BSs \cite{Palenzuela_2017,Bezares_2022, Helfer_2022, Sanchis_Gual_2022_proca, Evstafyeva_2023, Siemonsen_2023, Ge:2024itl} or a black hole (BH) and a BS \cite{Cardoso_2022, Zhong_2023}; some of these signals may be degenerate with those of standard BH binary mergers \cite{Evstafyeva:2024qvp, Sennett_2017, Bustillo_2021}. They have also been proposed as candidate objects for dark matter haloes \cite{Sin_1994, Schive_2014, Mourelle_24, Mourelle_25}, as black-hole mimickers capable of imitating a wide range of standard BH phenomenology \cite{Synge:1966okc,Cardoso:2008bp,Koga:2022dsu,Volkel:2022khh, Rosa_2022_mim, Rosa_2023}, and even as alternatives to central galactic BHs \cite{Torres:2000dw, Guzman:2005bs, Amaro_Seoane_2010, Rosa_2023, Rosa_2025}, though such alternatives are significantly constrained by observations made using the Event Horizon Telescope \cite{Olivares_2020}.

In addition to these astrophysical motivations, however, BSs are interesting for a wide range of numerical applications. These are enabled by their status as horizonless, localized clusters of energy that are -- crucially -- dynamically stable, on at least some branches of the parameter space \cite{Lee_1989}. One notable example of such a recent application involves numerically testing a conjectured instability of ultracompact objects related to the presence of a stable light ring \cite{Cunha:2017qtt,Cunha:2020azh, Keir:2014oka}; see Refs. \cite{Cunha:2022gde, Marks:2025, Staelens:2025wom, Evstafyeva:2025mvx}. There are others, however, that have a more mathematical character. For example, BSs have been used to study critical phenomena in spherical symmetry \cite{Lai_2007} and axisymmetry \cite{Lai_2004}, as well as to test the hoop conjecture via ultrarelativistic collisions \cite{Choptuik_2010}. 

{The study of higher-dimensional spacetimes dates back at least to efforts by Kaluza and Klein to unify gravity with electromagnetism by introducing an extra spatial dimension. It has since received attention in contexts ranging from string theory to gauge/gravity duality. More recently, the class of TeV-scale gravity theories \cite{Antoniadis:1990ew, Antoniadis_1998, Arkani_Hamed_1998} has motivated the numerical study of strong gravity in higher dimensions \cite{Witek_2011, Cook_2017, Sperhake_2019} to help search for observational signatures from extra dimensions. One may also consider such spacetimes from a mathematical perspective, however; see e.g. Ref.~\cite{Emparan_2008} for a review. This has been done in the context of critical phenomena, where the spherical collapse of gravitational waves was shown to be possible in $D = 5$ dimensions \cite{Bizo_2005}. A similar system incorporating a massless scalar field was later studied to help settle more general issues related to competition between vacuum and matter degrees of freedom in critical collapse \cite{Porto_Veronese_2022}. One might expect BSs could extend the understanding of higher-dimensional critical phenomena in at least two respects: by providing an example of Type I phenomena (as they have with $D = 4$ \cite{Hawley_2000}), and by presenting a natural way to study critical phenomena beyond spherical symmetry by critically tuning BS collisions. Similarly, implications of the hoop conjecture in $D = 5$ have been connected to the formation of naked singularities \cite{Yoo:2005nj}. Higher-dimensional BS collisions may thus be interesting not only as a test of the hoop conjecture for $D > 4$, but also as a potential mechanism for generating naked singularities.}

An issue with {these additional motivations} appears to manifest itself when we turn to the question of dynamical stability. While in four spacetime dimensions the gravitational attraction can balance the kinetic energy of the scalar field, one might expect that, because of the characteristic $1 / r^{D - 3}$ decay of the gravitational field, such a balance in $D > 4$ dimensions may be impossible, {causing BS solutions to be generically unstable.}
Previous work involving higher-dimensional models has included the construction of boson (spin-0), Dirac (spin-$1/2$) and Proca (spin-1) stars in $D \ge 4$ dimensions \cite{Blazquez-Salcedo:2019qrz}; this work did not however address the question of stability. A study of rotating mini BSs (BSs without self-interaction) in $D = 5$ was carried out in Ref. \cite{Brihaye_2016}, although this work suggested these models were unstable from the positive sign of the binding energy alone, which, although a useful clue, is in general neither a necessary nor a sufficient condition for instability-- see e.g. Ref. \cite{Marks_2025_CP}. More recently, the perturbative instability of spherically symmetric mini BSs in $D > 4$ dimensions has been rigorously shown by an analysis of radial oscillation frequencies \cite{Franzin_2024}. Higher-dimensional BSs have also been constructed and studied in asymptotically anti-de Sitter spacetimes \cite{Astefanesei_2003}.

It is thus a striking fact that the literature, to the best of our knowledge, presently contains no examples of higher-dimensional, asymptotically flat BS solutions that possess convincing evidence of dynamical stability. We note that, while a linear analysis is an important step towards determining the stability of a BS solution, it it not necessarily sufficient to draw a final conclusion. For example, rotating \cite{Di_Giovanni_2020} and excited \cite{Brito_2025} BSs that are stable to radial perturbations have been found to nonetheless suffer nonlinear instabilities when dynamically evolved. Conversely, it is possible in principle for nonlinear effects to re-stabilize a solution even in the presence of an unstable mode; this has been observed in the case of neutron-star models \cite{Sperhake_2001n}.

The purpose of this work is to extend the body of knowledge concerning BS solutions in $D \ge 4$ spacetime dimensions, considering both quartic and solitonic potentials. In particular, we demonstrate that either can be used to construct higher-dimensional BSs that are stable to spherical dynamics. We show this using both a perturbative analysis and a selection of dynamical evolutions in spherical symmetry. Our numerical approach to dynamical evolutions is based on the so-called ``modified cartoon'' method first introduced in Ref.~\cite{Pretorius_2004}, adapted to employ the Baumgarte-Shapiro-Shibata-Nakamura (BSSN) \cite{Baumgarte:1998te,Shibata:1995we} and conformal covariant Z4 (CCZ4) \cite{Alic:2011gg} formulations of numerical relativity (NR) in the presence of an $SO(D - 1)$ symmetry. We therefore retain the full freedom to choose gauge conditions supported by  standard contemporary approaches to NR, as well as enabling a straightforward comparison to the outputs of codes that use two- or three-dimensional grids.

Our paper is structured as follows. In Section~\ref{theory}, we review the theory under consideration and construct BS models in $D \ge 4$ dimensions, commenting on physical and asymptotic properties of the solutions. Following this, in Section~\ref{perturbative} we carry out a perturbative analysis of radial modes. We first derive the pulsation equations governing the linear dynamics of spherical perturbations with an arbitrary potential and in an arbitrary number of dimensions-- to our knowledge, the most general form of these equations present in the literature. We then apply these equations to solve for fundamental and first excited oscillation frequencies, thus demonstrating the existence of linearly stable branches in higher dimensions. Finally, in Section~\ref{dynamical} we report the results of a wide range of unperturbed and perturbed dynamical evolutions, showing that in all cases the results agree with the prediction of linear theory. We also use these evolutions to verify the robustness of our oscillation frequencies, and provide instability timescales for a range of perturbatively unstable models. Throughout, we use natural units setting $G = c = \hbar = 1$, but keep the scalar mass parameter $\mu$ under whose rescaling our BS solutions respect an appropriate scale symmetry.

\section{Theory and Model Construction} \label{theory}
We consider general relativity in $D$ spacetime dimensions, minimally coupled to a complex scalar field $\varphi$ with arbitrary potential $V(|\varphi|^2)$. Our action is thus
\begin{equation}
    \mathcal{S} = \int \du ^Dx\sqrt{-g}\left\{\frac{R}{16 \pi}  -\frac{1}{2}\left[
  g^{\mu\nu}\nabla_{\mu}\bar{\varphi}\,\nabla_{\nu}\varphi
  +V(|\varphi|^2)\right] \right\} \label{eq:action}
\end{equation}
where $R$ is the Ricci scalar, $g_{\mu\nu}$ is the spacetime metric, and an overbar denotes complex conjugation. Varying $\mathcal{S}$, we obtain the standard Einstein-Klein-Gordon equations:

\begin{eqnarray}
  && R_{\alpha\beta} - \frac{1}{2}R g_{\mu\nu}=8\pi \,T_{\alpha\beta}\,,
  ~~~~~
  \nabla^{\mu}\nabla_{\mu}\varphi = \frac{\du V}{\du \bar{\varphi}}\,, \label{eq:EKG}
  \\[5pt]
  && T_{\mu\nu}
  =
  \frac{1}{2} \nabla_{(\mu} \bar{\varphi}\,\nabla_{\nu)}\varphi
  -\frac{1}{2}g_{\mu\nu}
  \left[
  g^{\alpha\beta}\nabla_{\alpha}\bar{\varphi}\,\nabla_{\beta}\varphi
  + V(\varphi)
  \right].
  \nonumber
\end{eqnarray}
We now take the spherically symmetric metric ansatz,
\begin{equation}
    \du s^2 = -\alpha^2(r)\du t^2 + X^2(r)\du r^2 + r^2 \du \Omega^2_{D - 2},
\end{equation}
where 
\begin{equation}
    \du \Omega^2_{D - 2} = \sum_{i=1}^{D - 2}P(i)\du\theta_i^2, \quad P(i) = 
\begin{cases} 
      1 & i = 1\\
      \prod_{j=1}^{i-1} \sin^2\theta_j & i > 1 \\
   \end{cases} 
\end{equation}
is the usual metric for the $(D - 2)$-sphere. Taking also the harmonic ansatz $\varphi(t, r) = A(r)e^{i\omega t}$ for a spherical BS solution, we obtain the following system of ordinary differential equations (ODEs),
\begin{align}
   A'' &= \left(  \frac{X'}{X} - \frac{\alpha ' }{\alpha} - \frac{D - 2}{r}\right) + X^2\left(\frac{\du V}{\du |\varphi|^2} - \frac{\omega^2}{\alpha^2}\right)A, \label{eq:A_eq}\\ 
   \frac{X'}{X} &= -\frac{(D - 3)(X^2 - 1)}{2r} + \frac{4\pi r}{D - 2}\left[\frac{(A')^2}{X^2} + \frac{\omega^2A^2}{\alpha^2} + V(A^2) \right], \label{eq:X_eq}\\
   \frac{\alpha '}{\alpha} &= \frac{(D - 3)(X^2 - 1)}{2r} + \frac{4\pi r}{D - 2}\left[\frac{(A')^2}{X^2} + \frac{\omega^2A^2}{\alpha^2} - V(A^2) \right]. \label{eq:alpha_eq}
\end{align}
We solve this system using a shooting method to obtain asymptotically flat spacetimes, determining the appropriate BS frequency $\omega$ for each value of the central field amplitude $A_0.$ {The boundary conditions for this system are as follows,
\begin{align}
    A(0) &= A_0, \quad A'(0) = 0, \quad X(0) = 1,  \quad \alpha_0(\infty) = 1,  \quad A(\infty) = 1,
\end{align}
where the final condition is used to shoot for $\omega$.} Here, we limit our attention to BSs in the ground state, for which the central amplitude profile $A(r)$ has no zero crossings.

In this work, we focus on two choices of scalar potential. The first adds a quartic self-interaction term to the standard Klein-Gordon potential,
\begin{equation}
    V_\mathrm{massive}(|\varphi|^2) = \mu^2 |\varphi|^2 + \frac{\lambda}{2}|\varphi|^4
\end{equation}
for $\lambda$ constant. The resulting models are sometimes referred to as \textit{massive} BSs, while the special case $\lambda = 0$ corresponds to \textit{mini} BSs. We take $\lambda \ge 0$, corresponding to a repulsive self-interaction. We will generally present results in terms of the dimensionless rescaling $\hat\lambda := \lambda / \mu^2$. We also consider the so-called \textit{solitonic} potential, which has the form
\begin{equation}
    V_\mathrm{solitonic}(|\varphi|^2)= \mu^2 |\varphi|^2 \left(
  1-2\frac{|\varphi|^2}{\sigma_0^2}
  \right)^2.
\end{equation}
for $\sigma_0$ constant. Notice that the solitonic potential supports multiple vacuum states, at $|\varphi| = 0$ and additionally at $|\varphi| = \sigma_0 / \sqrt{2}$. The presence of this degenerate vacuum is known in $D = 4$ to significantly raise the compactnesses achievable in the stable regime \cite{Lee:1987, Collodel:2022jly}. With the potential chosen, our action~\eqref{eq:action} is invariant under a rescaling in terms of the dimensionless quantities,
\begin{equation}
    x^\nu \rightarrow \mu x^\nu, \quad \omega \rightarrow \frac{\omega}{\mu}, \quad M \rightarrow \mu^{D - 3}M, \quad \lambda \rightarrow \frac{\lambda}{\mu^2}.
\end{equation}
We are therefore able to set $\mu = 1$ in our numerical solver without loss of generality.

Before moving on, we note that to compute the numerically most challenging models, we will employ a two-way shooting code, similar to that used in Ref.~\cite{Evstafyeva:2023kfg}. To apply this method effectively, we will require the asymptotic behavior of our solutions, before making use of an alternative decomposition in the new variable $y := 1/r$ in the BS exterior. We first introduce the function
\begin{equation}
    m(r) := \frac{r^{D-3}}{2}\left(\frac{X^2 - 1}{X^2}\right)
\end{equation}
so that $X = \left[1 - 2m(r)/r^{D-3}\right]^{-1/2}$. Note, however, that for $D > 4$, the asymptotic value of $m(r)$ does \textit{not} yield the total Arnowitt–Deser–Misner (ADM) mass $M$ of the spacetime; the two are instead related by a geometric factor,
\begin{equation}
    M = \frac{(D - 2)\Omega_{D - 2}}{8\pi}\lim_{r\to\infty}m(r),
\end{equation}
where $\Omega_{D - 2} = 2\pi^{(D - 1)/2} / \Gamma\left(\frac{D-1}{2}\right)$ is the surface area of the unit $(D - 2)$-sphere. We also note that we can always fix the gauge so that $\alpha\rightarrow 1$ at spatial infinity by rescaling $t$; from now on we make this choice.

To obtain a set of equations for the exterior region, we first note the BS asymptotics at large radii. For $D = 4$, a careful asymptotic expansion of Eq.~\eqref{eq:A_eq} reveals that the solution must, at leading order, take the form {(see Ref.~\cite{Evstafyeva:2023kfg} for more details)}
\begin{equation}
    A(r) \sim \frac{ e^{-\sqrt{1 - \omega^2}r}}{r^{1 + \epsilon}}, \quad \epsilon = \frac{1 - 2\omega^2}{\sqrt{1 - \omega^2}}M \quad (D = 4).
\end{equation}
However, for $D > 4$ {performing the same asymptotic expansion shows that the leading-order term is independent of $M$ and $\omega$,} and we are left with the simpler asymptotic behavior,
\begin{equation}
    A(r) \sim \frac{ e^{-\sqrt{1 - \omega^2 }r}}{r^{\frac{D - 2}{2} }} \quad (D > 4).
\end{equation}
Thus we see that BSs in $D = 4$ are an asymptotically special case. Because of this exponential decay, BSs do not have a hard surface. We therefore define an effective radius $r_{99}$ as that enclosing $99 \%$ of the mass. Defining $k := \sqrt{1 - \omega^2}$ and introducing the new quantities $\xi:= A e^{k / y}, \quad \kappa := -A' e^{k / y}/X,$ a brief calculation allows us to write a first-order system appropriate for the BS exterior,
\begin{align}
  \xi' &= \frac{X \kappa - k\xi}{y^2}, \\ 
    m' &=  -\frac{4\pi}{(D - 2)y^D}\left[ \left(\kappa^2 + \frac{\omega^2\xi^2}{\alpha^2}\right)e^{-2k/y} + V(A^2) \right],\\
   \frac{\alpha'}{\alpha} &= \frac{(D - 3)m}{1 - 2my^{D-3}} - \frac{4\pi}{(D - 2)y^3(1 - 2my^{D - 3})}\left[ \left(\kappa^2 + \frac{\omega^2\xi^2}{\alpha^2}\right)e^{-2k/y} - V(A^2) \right], \\
   \kappa' &= \frac{(D - 2)\kappa}{y} - \kappa\frac{ \alpha'}{\alpha} - - \frac{k\kappa}{y^2} + \frac{X\xi}{y^2}\left[\frac{\du V}{\du |\varphi|^2}  -\frac{\omega^2}{\alpha^2}   \right],
\end{align}
where primes now represent derivatives with respect to $y$. {This system is used along with Eqs.\eqref{eq:A_eq}--\eqref{eq:alpha_eq} in our two-way shooting code, and is especially important for computing very compact BS solutions accurately.}

We also note at this stage that the $U(1)$ symmetry of the action \eqref{eq:action} is associated with a conserved current $J^\mu = \frac{1}{2}i(\varphi\nabla^\mu\bar\varphi - \bar\varphi\nabla^\mu \varphi).$ This gives rise to a conserved Noether charge, which one can interpret as a particle number,

\begin{equation} \label{eq:noether}
    N = \int \du ^ D x \;\sqrt{-g} \;J^0 = \omega\Omega_{D - 2} \int \du r \; r^{D - 2} \frac{X}{\alpha}A^2.
\end{equation}
The \textit{binding energy} is then defined as $E_B := M - \mu N$, and measures the work required to assemble the BS from spacelike infinity. Solutions with negative $E_B$ are thus interpreted as gravitationally bound; those with positive $E_B$ are interpreted as gravitationally unbound.

\begin{figure}
 \centering
        \includegraphics[width=\linewidth]{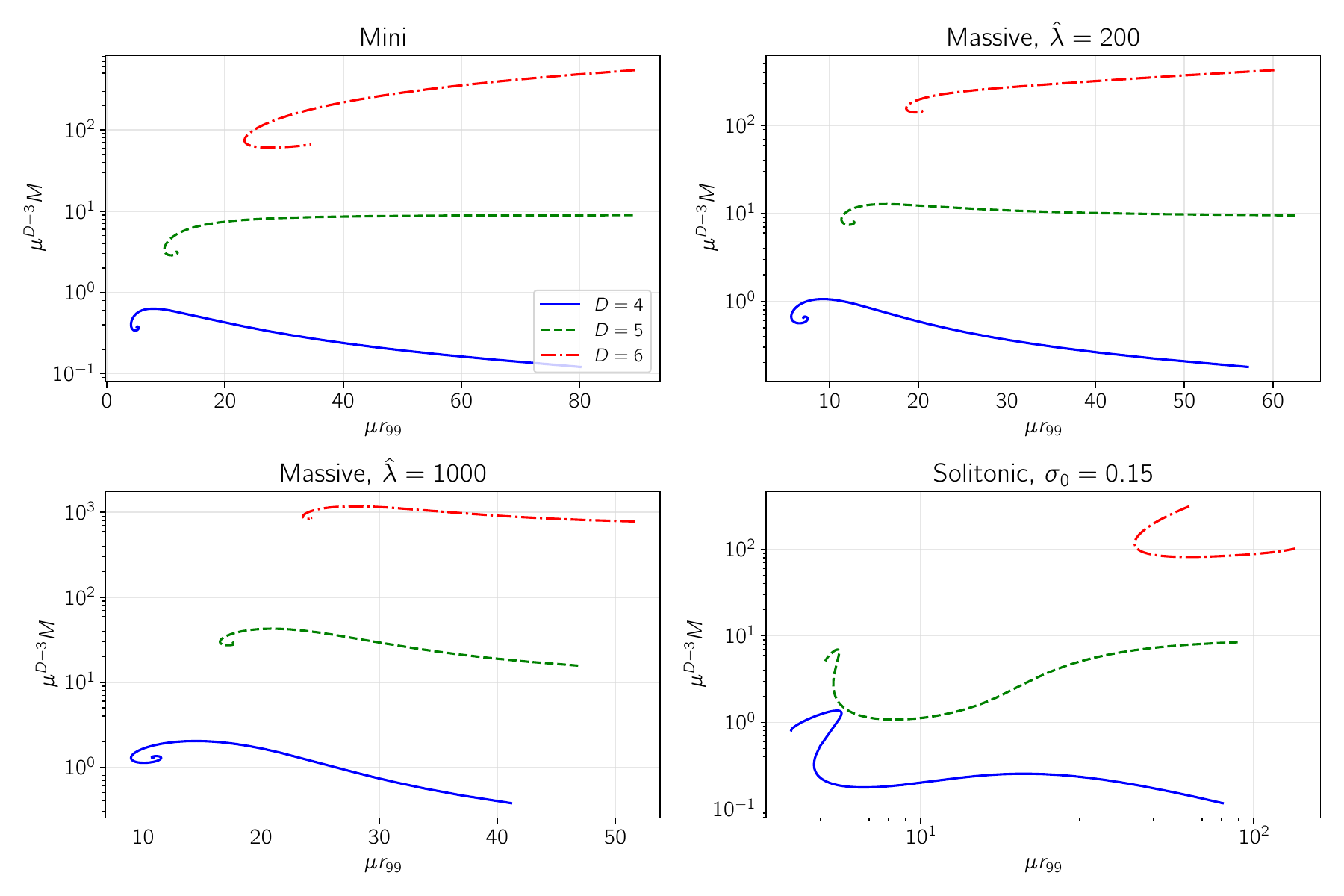}
 \caption{Mass against radius curves for a selection of BS families, showing the impact of the spacetime dimension. For massive models, $\hat\lambda := \lambda / \mu^2 .$}
\label{fig:M_vs_R}
\end{figure}

\begin{figure}
 \centering
        \includegraphics[width=\linewidth]{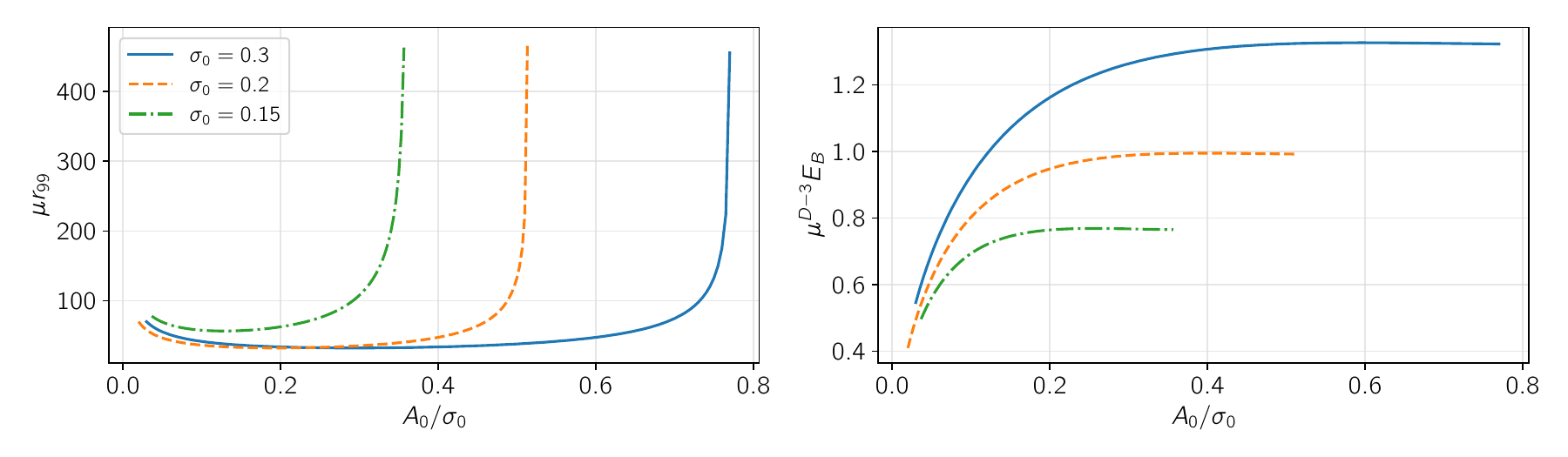}
 \caption{Radius (left) and binding energy (right) against amplitude for solitonic families in $D = 6$, showing the divergence in $r_{99}$ at sufficiently large central amplitude even as the binding energy approaches a limiting value. }
\label{fig:sigma_divergence}
\end{figure}

In Fig.~\ref{fig:M_vs_R} we plot the BS mass against radius $r_{99}$ for a selection of families. We note a few interesting properties of the solution spaces at this stage. First, as seen in previous studies of higher-dimensional mini BSs \cite{Blazquez-Salcedo:2019qrz, Franzin_2024}, we find that for $D > 4$ in all cases the limit of central amplitude $A_0 \rightarrow 0$ does not correspond also to $M \rightarrow 0$. Rather, there is a finite mass gap between the vacuum Minkowski ground state and the space of BS solutions with nonzero boson number. In addition, we note a feature seen only in solitonic models in $D = 6$ with sufficiently small $\sigma_0$. Generally, for large $A_0$ the mass $M$, radius $r_{99}$, and frequency $\omega$ all approach finite values corresponding to a limiting solution. We see this behavior borne out in all cases in $D = 4$ and $5$ dimensions. However, for solitonic models in $D = 6$ with $\sigma_0 \lesssim 0.35$, we see that $M$ and $r_{99}$ diverge, although the binding energy $E_B$ approaches a constant. This behaviour is shown more clearly in Fig.~\ref{fig:sigma_divergence}. {We cannot construct any models with central amplitude larger than that at which this apparent divergence occurs.  While we cannot rule out the possibility that improved numerics will find them, we conjecture that they do not exist. The divergence may be due to the leading-order behavior of the solitonic potential, which is that of an attractive self-interaction. This could possibly overwhelm the ability of the field to balance the combined gravitational and scalar attraction in more than five dimensions, but we leave a detailed mathematical investigation to future study.}

\begin{figure} 
 \centering
        \includegraphics[width=\linewidth]{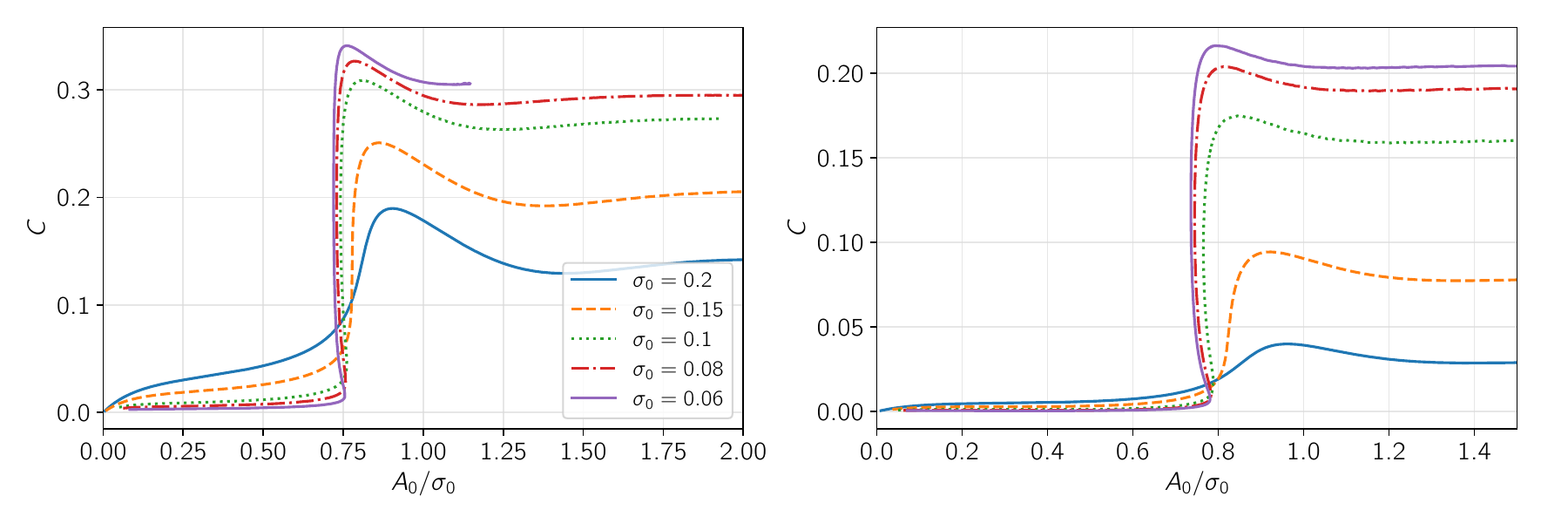}
 \caption{Compactness (defined in Eq.~\eqref{eq:C_def}) against central amplitude for solitonic families in $D = 4$ (left) and $5$ (right) dimensions. }
\label{fig:sigma_compactness}
\end{figure}

For comparison, the effect of lowering the solitonic parameter $\sigma_0$ in $D = 5$ much more closely resembles that seen in $D = 4$, causing the family of solutions to become multi-valued as a function of $A_0$ and allowing one to achieve very high compactness 
\footnote{One might worry that the apparent divergence we see in the solution space for $D = 6$ reflects that we are actually entering the multivalued regime, and are simply failing to compute the higher-mass models. We do not believe this to be the case for two main reasons. Firstly, in $D = 4, 5$, the multivalued regime begins for each family close to the degenerate vacuum at $A_0 \sim \sigma_0 / \sqrt{2}$, while in $D = 6$ the divergence occurs at varied, and generally much smaller, central amplitudes. Secondly, in $D = 4, 5$ we see that $\frac{\du E_B}{\du A_0}$ also approaches infinity where $\frac{\du M}{\du A_0}$ does, but in $D = 6$ it approaches zero.}. 
To illustrate this, we define the compactness as
\begin{equation}
    C := \frac{m(r_{99})}{r_{99}^{D-3}} \label{eq:C_def}
\end{equation}
and plot it against $A_0$ in $D = 4,5$ for a range of $\sigma_0$ values in Fig.~\ref{fig:sigma_compactness}. As a reference, the location of the light ring for a Tangherlini black hole would correspond to $C = \frac{1}{D - 1}$. Indeed, we can verify that, unlike in $D = 4$~\cite{Collodel:2022jly}, all the models we have computed in $D = 5$ do not support light rings. Results previously obtained using the thin-wall approximation, in which one prescribes a simple Heaviside-function profile for the field amplitude and solves the metric equations \eqref{eq:X_eq}, \eqref{eq:alpha_eq}, suggest that models with a pair of light rings may however exist for sufficiently small $\sigma_0;$ cf. Ref.~\cite{Cardoso_2022_eco}.

\section{Perturbative Analysis} \label{perturbative}
In this section, we conduct a perturbative analysis of the higher-dimensional BS solutions presented in the previous section. Our approach is based primarily on methods first introduced in Refs.~\cite{Gleiser:1988ih, Gleiser:1988rq}, and generalizes formalisms used to compute oscilation frequencies in Refs.~\cite{Hawley_2000, Kain:2021rmk, Franzin_2024, Santos:2024vdm}. Before we begin this analysis, however, it is helpful to first consider briefly the binding energy of the solutions we have considered thus far. Plotting the binding energy against Noether charge for our solution families, we obtain what are called \textit{bifurcation diagrams}. In addition to the hint provided by the sign of the binding energy alone, arguments from catastrophe theory suggest that information about perturbative stability can be inferred by interpreting the cusp structure of the bifurcation diagram \cite{Kusmartsev:1990cr}. In particular, when moving from a lower to a higher branch we expect a stable mode to become unstable, and vice versa. A limitation of this approach is that it cannot necessarily determine the number of unstable modes present on the first branch of the solution space, meaning that the overall picture of stability may be ambiguous without a perturbative analysis. 

\begin{figure}
 \centering
        \includegraphics[width=\linewidth]{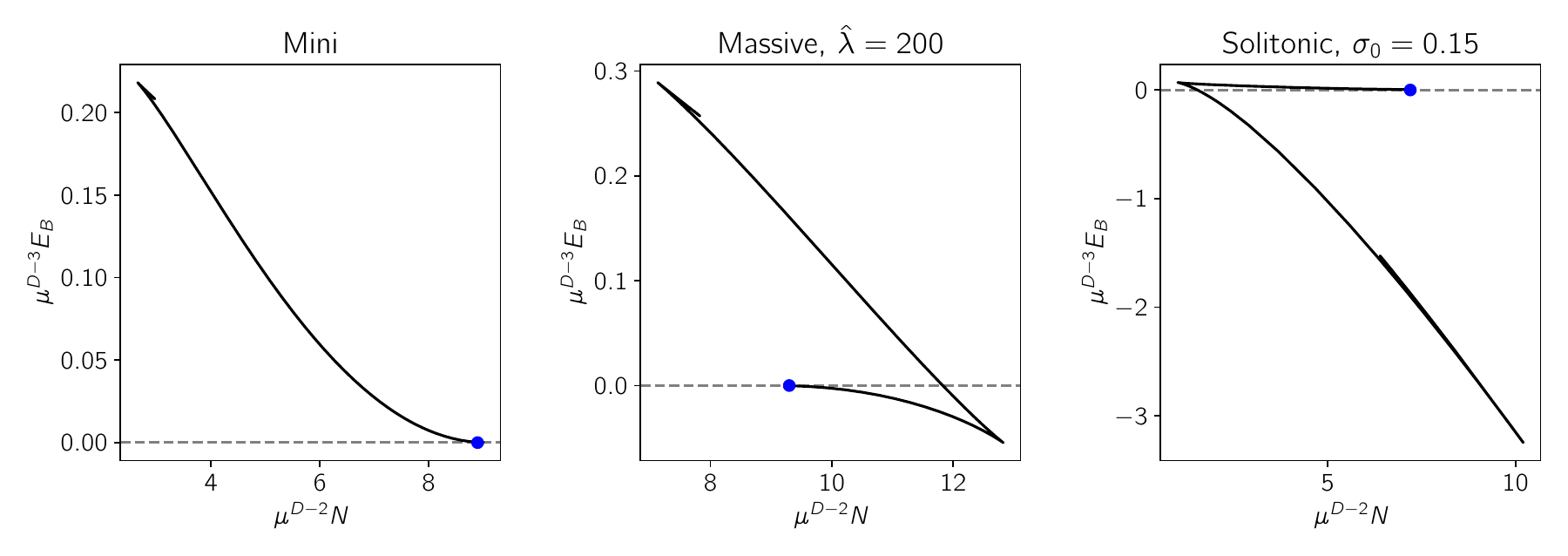}
 \caption{Bifurcation diagrams for representative mini, massive, and solitonic solution families in $D = 5$ spacetime dimensions. As a visual aid, we indicate the limiting solution as $A_0 \rightarrow 0$ with a blue dot.}
\label{fig:bifurcation_diagrams}
\end{figure}

With these caveats made, we show bifurcation diagrams for a selection of families in $D = 5$ dimensions. For mini BSs, the binding energy is everywhere positive, strongly suggesting that there are no stable mini BS models with $D = 5$, as we know to be true. However, for the other two families we see branches of the solution space with mostly or entirely negative binding energy. This, combined with the cusp structure, suggests that the first massive BS branch and second solitonic BS branch could house perturbatively stable higher-dimensional models for appropriate values of $\hat\lambda$ and $\sigma_0$. This would make sense in light of the previous observation that sufficiently strong self-interactions can stabilize rotating \cite{Siemonsen_2021} and excited \cite{Sanchis_Gual_2022} BS models in $D = 4$ that would otherwise be unstable. We will now show that this is indeed the case.

\subsection{The Pulsation Equations}
{Here, we follow the approach presented in Appendix III of Ref.~\cite{Marks:2025}, generalizing the results of that work.}
In this section, we use overdots to denote time derivatives and primes to denote radial derivatives. We first decompose the complex scalar field into two real fields, $\varphi  = e^{i \omega t} \left( \psi_1 + i\psi_2 \right)\,$. Next, we introduce the perturbation functions $\delta \psi_1$,
$\delta \psi_2$, $\delta \alpha$ and $\delta X$ around a stationary BS background according to
\begin{align}
    \psi_1  = A_\mathrm{b}(1 + \delta\psi_1), \quad \psi_2   =A_\mathrm{b} \delta \psi_2\,,  \quad
    \alpha  = \alpha_\mathrm{b}\left(1 + \frac 12 \delta \alpha \right), \quad \ X = X_\mathrm{b}\left(1 + \frac 12 \delta X \right). \label {eq:perts}
\end{align}
where $\{A_\mathrm{b}, X_\mathrm{b}, \alpha_\mathrm{b} \}$ solves the background equations~\eqref{eq:A_eq}--\eqref{eq:alpha_eq}. Substituting these into the EKG system \eqref{eq:EKG} and linearizing, we obtain a system of four ODEs in the perturbation functions. Introducing $n := D - 2$ for convenience, we can write these as follows:
\begin{align}
    \label{eq:psi1_rr}\delta\psi_1'' = \, &\frac{X_\mathrm{b}^2}{\alpha_\mathrm{b}^2}\left(\delta\ddot\psi_1+ 2\omega\dot\psi_2 + \omega^2\delta \alpha \right) + X_\mathrm{b}^2\left(\frac{\du V}{\du |\varphi|^2} - \frac{\omega^2}{\alpha_\mathrm{b}^2} \right)\delta X + 2A_\mathrm{b}^2X_\mathrm{b}^2\frac{\du^2 V}{\du (|\varphi|^2)^2}\delta\psi_1 \\ &\;+ \left(\frac{X_\mathrm{b}'}{X_\mathrm{b}} - \frac{\alpha_\mathrm{b}'}{\alpha_\mathrm{b}} - \frac{n}{r} \right)\delta\psi_1' + \frac{A_\mathrm{b}'}{2A_\mathrm{b}}\left(\delta X' - \delta \alpha' - 4\delta\psi_1' \right), \nonumber
    \\
     \label{eq:lambda_r}\delta X' = \, & \frac{16\pi r A_\mathrm{b}^2 X_\mathrm{b}^2}{n\alpha_\mathrm{b}^2}  \left(\omega^2\delta \alpha - \omega\delta\dot\psi_2 + \frac{A_\mathrm{b}'\alpha_\mathrm{b}^2}{A_\mathrm{b}}\delta\psi_1' + \left[\frac{\omega^2}{X_\mathrm{b}^2} + \alpha_\mathrm{b}^2\frac{\du V}{\du |\varphi|^2} + \frac{(A_\mathrm{b}')^2\alpha_\mathrm{b}^2}{A_\mathrm{b}^2} \right]\delta\psi_1 \right)\\
     &+ \left(\frac{2X_\mathrm{b}'}{X_\mathrm{b}} - \frac{8\pi r(A_\mathrm{b}')^2}{n} - \frac{D - 3}{r} \right)\delta X,
     \nonumber 
     \end{align}
\begin{align}
     \label{eq:lambda_tt}\delta\ddot X = \, & \frac{2\alpha_\mathrm{b}}{X_\mathrm{b}^2}\left( \frac{(D - 3)\alpha_\mathrm{b}X_\mathrm{b}'}{rX_\mathrm{b}} - \alpha_\mathrm{b}'' + \left[\frac{X_\mathrm{b}'}{X_\mathrm{b}} - 1\right]\alpha_\mathrm{b}' - \frac{8\pi(D-3)\alpha_\mathrm{b}(A_\mathrm{b}')^2}{n} \right)\delta X + \frac{\alpha_\mathrm{b}^2}{X_\mathrm{b}^2}\delta \alpha'' \\ \nonumber
     &+ \frac{32 \pi \alpha_\mathrm{b}^2}{nX_\mathrm{b}^2}\left[\left(A_\mathrm{b}^2\frac{\du V}{\du |\varphi|^2} - \omega^2A_\mathrm{b}^2+ \frac{D - 3}{n}(A_\mathrm{b}')^2 \right)\delta\psi_1 + \frac{D - 3}{n}A_\mathrm{b}'A_\mathrm{b}\delta\psi_1' \right] 
     \\ \nonumber &+ \frac{\alpha_\mathrm{b}^2}{X_\mathrm{b}^2}\left(\frac{2\alpha_\mathrm{b}'}{\alpha_\mathrm{b}} - \frac{X_\mathrm{b}'}{X_\mathrm{b}} + \frac{1}{r}\right)\delta \alpha' +  \frac{16\pi \omega A_\mathrm{b}^2}{n}\left(\omega\delta \alpha+ 2\delta\dot\psi_2\right) - \frac{\alpha_\mathrm{b}^2}{X_\mathrm{b}^2}\left(\frac{\alpha_\mathrm{b}'}{\alpha_\mathrm{b}} + \frac{D - 3}{r}\right)\delta X',
     \nonumber 
     \\
     \label{eq:nu_r}\delta X' - \delta \alpha' = \, & 2\left( \frac{X_\mathrm{b}'}{X_\mathrm{b}} - \frac{\alpha_\mathrm{b}'}{\alpha_\mathrm{b}} - \frac{D - 3}{r} \right)\delta X + \frac{32\pi rA_\mathrm{b}^2X_\mathrm{b}^2}{n}\frac{\du V}{\du |\varphi|^2}\delta\psi_1.
\end{align}
We now take the harmonic ansatz $\delta\psi_1(t,r) = e^{\iu \chi t}f(r),$   $\delta X(t,r) = e^{\iu \chi t} g(r) $, leaving $\chi\in \mathbb{C}$ undetermined. The resulting two equations form a self-adjoint system.  Therefore, like the BS frequencies $\omega$, the radial oscillation frequencies $\chi$ are ordered in an infinite discrete sequence $\chi_0^2 < \chi_1^2 < ...$, where the solution with eigenvalue $\chi_n^2$ is regular with $n$ zero crossings in $f$ and $g$. A negative $\chi_0^2$ thus signifies a radial instability, while a positive $\chi_0^2$ guarantees that the model is radially stable at the linear level.

To obtain the first pulsation equation, we solve Eq.~\eqref{eq:lambda_r} for $\delta\dot\psi_2$ and substitute the result into Eq.~\eqref{eq:psi1_rr}. This yields an equation that depends on $\delta \alpha$ only through its radial derivative, which can then be removed using Eq.~\eqref{eq:nu_r}. To derive the second
pulsation equation, we combine Eqs.~\eqref{eq:lambda_r} and~\eqref{eq:lambda_tt} with the background equations \eqref{eq:A_eq}--\eqref{eq:alpha_eq}, obtaining an equation that contains $\delta \alpha$ only through its first and second radial derivatives. By using Eq.~\eqref{eq:nu_r}
and its derivative, we can finally eliminate $\delta \alpha$ 
and arrive at our final system of ODEs, the pulsation equations in an arbitrary spacetime dimension and with an arbitrary potential:
\begin{equation}
\label{eq:f_eq}
    \begin{split}
        f'' = & \Bigg[ \left( \frac{2\omega^2 - \chi^2}{\alpha_\mathrm{b}^2} + 2A_\mathrm{b}^2 \frac{\du^2 V}{\du (|\varphi|^2)^2} + 2\frac{\du V}{\du |\varphi|^2} \right) + \frac{16\pi}{n} r A_\mathrm{b} A_\mathrm{b}' \frac{\du V}{\du |\varphi|^2} + \frac{2(A_\mathrm{b}')^2}{A_\mathrm{b}^2} \Bigg] X_\mathrm{b}^2 f \\
        & + \Bigg[ \left(-\frac{\omega^2}{\alpha_\mathrm{b}^2} + \frac{\du V}{\du |\varphi|^2}\right)X_\mathrm{b}^2 - \frac{n^2-n}{8\pi r^2 A_\mathrm{b}^2 } - \frac{A_\mathrm{b}'\alpha_\mathrm{b}'}{A_\mathrm{b}\alpha_\mathrm{b}} + \frac{(A_\mathrm{b}')^2}{A_\mathrm{b}^2} - (D - 3) \frac{A_\mathrm{b}'}{r A_\mathrm{b}} + \frac{A_\mathrm{b}'X_\mathrm{b}'}{A_\mathrm{b}X_\mathrm{b}} + \frac{nX_\mathrm{b}'}{4\pi r A_\mathrm{b}^2  X_\mathrm{b}} \Bigg] g \\
        & + \left( \frac{X_\mathrm{b}'}{X_\mathrm{b}} - \frac{n}{r} - \frac{\alpha_\mathrm{b}'}{\alpha_\mathrm{b}} \right) f' - \frac{n}{8\pi r A_\mathrm{b}^2 } g'
    \end{split}
    \end{equation}
\begin{equation}
\label{eq:g_eq}
    \begin{split}
        g'' = & 32 \pi\Bigg[\frac{2 r  }{n} \Bigg(
      A_\mathrm{b}' A_\mathrm{b}^3 \frac{\du^2 V}{\du (|\varphi|^2)^2}
    +  A_\mathrm{b} A_\mathrm{b}' \frac{\du V}{\du |\varphi|^2}
    + \frac{\alpha_\mathrm{b}'}{\alpha_\mathrm{b}}  A_\mathrm{b}^2  \frac{\du V}{\du |\varphi|^2}
  \Bigg) X_\mathrm{b}^2 + \frac{ r}{n} A_\mathrm{b}^2 X_\mathrm{b}  \frac{\du V}{\du |\varphi|^2} X_\mathrm{b}'
-  \left( A_\mathrm{b}' \right)^2 \Bigg] f \\
        & + \Bigg[ -\frac{\chi^2X_\mathrm{b}^2}{\alpha_\mathrm{b}^2} - \frac{4n-4}{r\alpha_\mathrm{b}}\alpha_\mathrm{b}' - \frac{2-2n}{r^2} + 16\pi(A_\mathrm{b}')^2 - \frac{2(\alpha_\mathrm{b}')^2}{\alpha_\mathrm{b}^2} + \Bigg(
    \frac{4X_\mathrm{b}' \alpha_\mathrm{b}'}{\alpha_\mathrm{b} X_\mathrm{b}} 
    + 2 X_\mathrm{b}''
    - \frac{2}{r} X_\mathrm{b}'
  \Bigg)  
  - \frac{4\left( X_\mathrm{b}' \right)^2}{X_\mathrm{b}^2}  \Bigg] g \\
        & + 32\pi\left(
    \frac{r}{n} A_\mathrm{b}^2 X_\mathrm{b}^2  \frac{\du V}{\du |\varphi|^2}
    -  A_\mathrm{b}' A_\mathrm{b}
\right) f' + \left( \frac{2 - n}{r} - \frac{3\alpha_\mathrm{b}'}{\alpha_\mathrm{b}} + \frac{3X_\mathrm{b}'}{X_0}  \right) g'
    \end{split}
    \end{equation}
To integrate out of the singularity at $r = 0$, we must determine the small-$r$ expansion of these equations. Using the even parity of scalar functions across $r=0$ and our freedom
to rescale the perturbation functions, we can expand them without loss of generality
as $f = 1 + \frac a 2 r^2 + \mathcal{O}(r^4)$ and $g = \beta + \frac \gamma 2 r^2 + \mathcal{O}(r^4)${, supplying boundary conditions for our solver at the origin}. Expanding Eq.~\eqref{eq:f_eq} in this manner leads to
  \begin{equation}
        a = \frac{2}{(n + 1) \alpha_0^2 }\left(\omega^2 - \frac{1}{2}\chi^2  + \frac{\alpha_0^2}{A_0} \frac{\du^2 V}{\du (|\varphi|^2)^2}(A_0^2) + \frac{\du V}{\du |\varphi|^2}(A_0^2) \alpha_0^2 - \frac{n(n+1)}{32\pi\gamma}\frac{\alpha_0^2}{A_0^2} \right),
    \end{equation}
where $A_0$ and $\alpha_0$ are the central amplitude and lapse. On the other hand, expanding Eq~\eqref{eq:g_eq} shows $\beta = 0$ but leads to a degeneracy at the following order, so that $\gamma$ remains free. We thus have {\it two} undetermined parameters, $\chi$ and $\gamma$, and must shoot for both to obtain asymptotically flat spacetimes. 

We finally note that, by using of the perturbation \eqref{eq:psi1_rr}--\eqref{eq:nu_r} and background \eqref{eq:A_eq}--\eqref{eq:alpha_eq} equations, one can carry out the integral describing the perturbation to the Noether charge \eqref{eq:noether} explicitly. The result is
\begin{equation}
    \delta N = \lim_{r \to \infty} \Omega_{D - 2}\left[
        \frac{2  \alpha_\mathrm{b} A_\mathrm{b} A_\mathrm{b}' r^n}{\omega\, X_\mathrm{b}} \, \delta\psi_1
        - \frac{n \alpha_\mathrm{b} r^{n-1}}{8\pi\, \omega\, X_\mathrm{b}} \, \delta X
        \right].
\end{equation}
The condition $\delta N = 0$ can either be {treated as an equation to be solved numerically via e.g. a secant method, or to help determine the accuracy of an oscillation frequency computed by counting the number of zero crossings}. Emprically, we find that when shooting for a frequency $\chi^2 > 0$, {we reliably obtain convergence} when counting the number of zero crossings and checking that $\delta N \ll N$ at the end. Conversely, when computing a frequency $\chi^2 < 0$, we {more reliably obtain convergence} when we apply a secant method to solve $\delta N = 0$, and then count the number of zero crossings to ensure we have found the desired eigenvalue. {This is because the profiles of $f$ and $g$ associated with negative $\chi^2$ values seem more prone to exhibiting spurious zero crossings when $\chi^2$ is not close to an eigenvalue.}

\subsection{Radial Oscillation Frequencies}
\begin{figure}
 \centering
        \includegraphics[width=\linewidth]{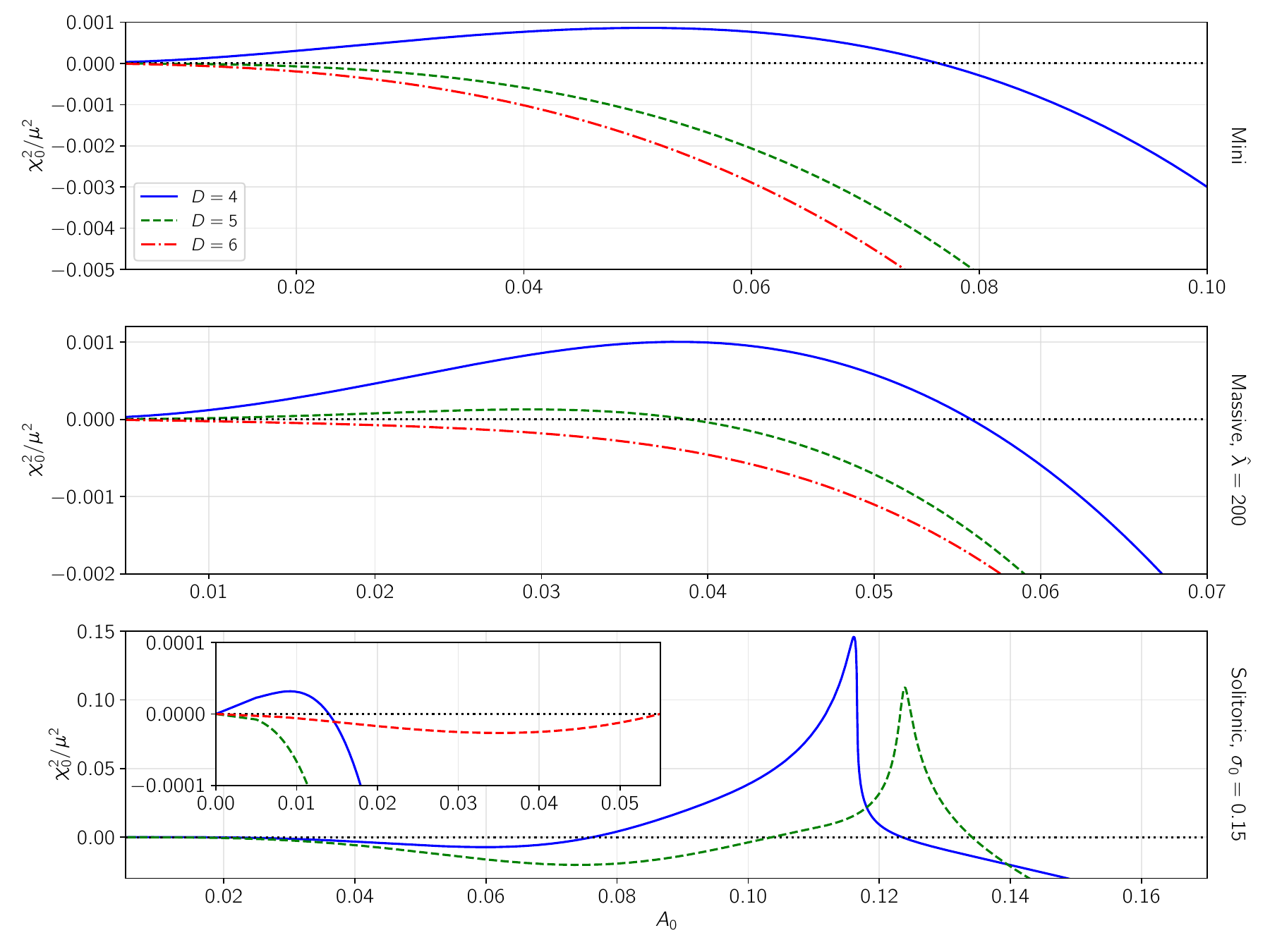}
 \caption{Fundamental radial oscillation frequency $\chi_0^2$ against central amplitude $A_0$ for mini, massive, and solitonic BS models in varying dimension. Regions with $\chi_0^2 > 0$ correspond to linear radial stability. Inset on the bottom panel shows the first stable branch in $D = 4$, which is absent in $D = 5$, as well as the result for $D = 6$. Notice $\chi_0^2 \rightarrow 0$ as the solution family diverges in the $D = 6$ case.}
\label{fig:ground_modes}
\end{figure}

\begin{figure}
 \centering
        \includegraphics[width=\linewidth]{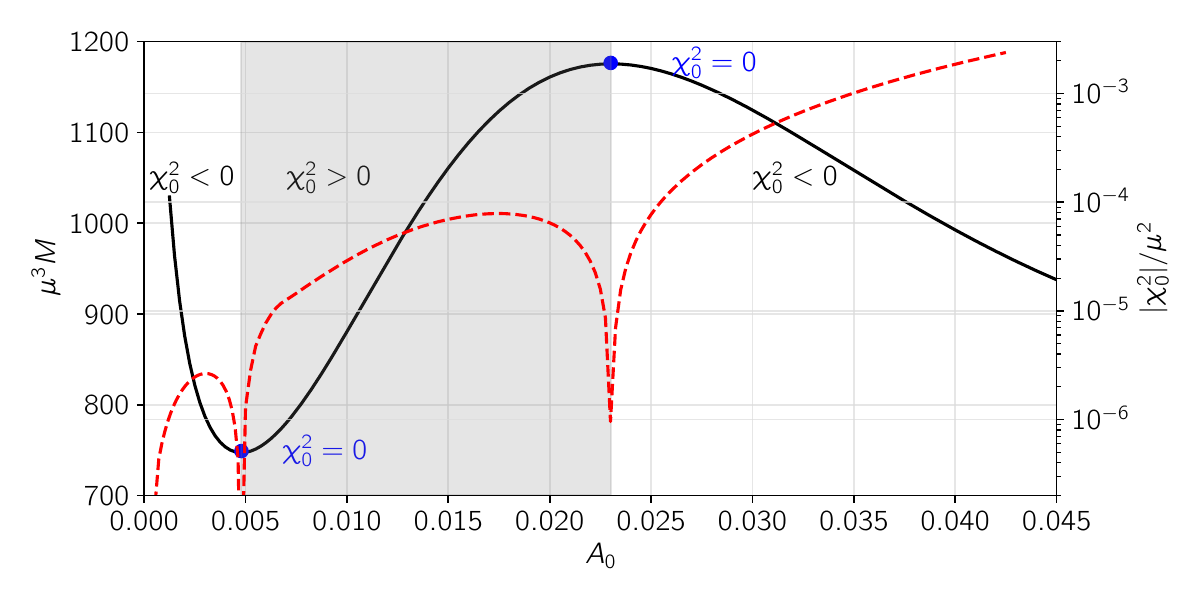}
 \caption{The solid black line shows the mass $M$, and the dashed red line the {absolute value of} the fundamental radial oscillation frequency $\chi_0^2$, both against the central amplitude $A_0$ for a massive BS family in $D = 6$ with $\hat\lambda = 1000$. The shaded region corresponds to perturbatively stable models, i.e. those with $\chi_0^2 > 0$. We draw attention to the fact that this region does not extend to $A_0 = 0$, unlike in $D = 4$ and $D = 5$. This feature is present for all massive families in $D = 6$ with $\hat\lambda$ larger than the critical value given in Table~\ref{tab:critical_values}.}
\label{fig:D6_ground}
\end{figure}

Here, we present oscillation frequencies computed using the formalism developed in the previous section. Our results are consistent with those presented in Refs.~\cite{Hawley_2000,Kain:2021rmk, Marks:2025} for mini, massive, and solitonic BSs in $D = 4$. In Fig.~\ref{fig:ground_modes} we display the fundamental mode $\chi_0^2$ for a sampling of families, showing the impact of the spacetime dimension. As expected, we see that stable mini BSs exist only in $D = 4$. However, as we increase the self-interaction strength $\hat\lambda$, perturbatively stable branches eventually emerge in $D = 5$ and $6$ {(see Fig.~\ref{fig:D6_ground} below for $D = 6$)}. Furthermore, for sufficiently small solitonic parameter $\sigma_0,$ a stable branch emerges in $D = 5$, corresponding to the so-called relativistic branch that emerges for sufficiently small $\sigma_0$ in $D = 4$. Unlike in $D = 4$, however, there is only ever one stable branch for solitonic models in $D = 5$. We also note that we are unable to find any stable solitonic models in $D = 6$, which is unsurprising given the divergences reported in Section \ref{theory} at low compactness. As an additional check, we find that the zero crossings of the ground mode always correspond to an extremum in the curve $M(A_0)$, as expected. Thus, our expectations based on the bifurcation diagrams shown in Fig.~\ref{fig:bifurcation_diagrams} are fully confirmed by a perturbative analysis.

In Table~\ref{tab:critical_values}, we display the critical values of $\hat\lambda$ and $\sigma_0$ needed to change the branch structure of the BS solution space from that corresponding to mini BSs. These are to be interpreted as follows. For massive models, a stable branch exists for $\hat\lambda$ larger than the critical value, connected to $A_0 = 0$. For solitonic models, a stable branch exists for $\sigma_0$ smaller than the critical value, corresponding to the region between the first minimum in the $M(A_0)$ curve and the following maximum, though in $D = 4$ there is an additional stable branch starting at $A_0 = 0$. No such $\sigma_0$ exists in $D = 6$.
\begin{table}
\caption{Critical values of $\hat\lambda$ and $\sigma_0$ required for the emergence of new stable branches in $D = 4,5,6$.}
\vspace{5pt}
\centering
\begin{tabular}{l c c c}
\hline
$D$ & 4 & 5 & 6 \\
\hline
Critical $\hat\lambda$ & 0 & 63.4 & 416 \\
Critical $\sigma_0$ & 0.322 & 0.236 & n/a \\
\hline
\end{tabular}
\label{tab:critical_values}
\end{table}

As indicated by Table~\ref{tab:critical_values}, for sufficiently large $\hat\lambda$ we are also able to construct perturbatively stable models in $D = 6$. Such families have an interesting feature, which we show in Fig.~\ref{fig:D6_ground}: there is an additional local minimum in the $M(A_0)$ curve which is not present in $D = 4,5,$ and which corresponds to a zero crossing of $\chi_0^2$. Thus, the stable branch of massive boson stars in six dimensions, unlike in the four- and five-dimensional cases, does not include the limiting solution at $A_0$ = 0. This also implies that stable six-dimensional models of arbitrarily low compactness cannot be found for a given quartic potential, although it appears one can make the range of unattainable compactnesses as small as desired by choosing a sufficiently large $\hat\lambda$. 

\begin{figure}
 \centering
        \includegraphics[width=\linewidth]{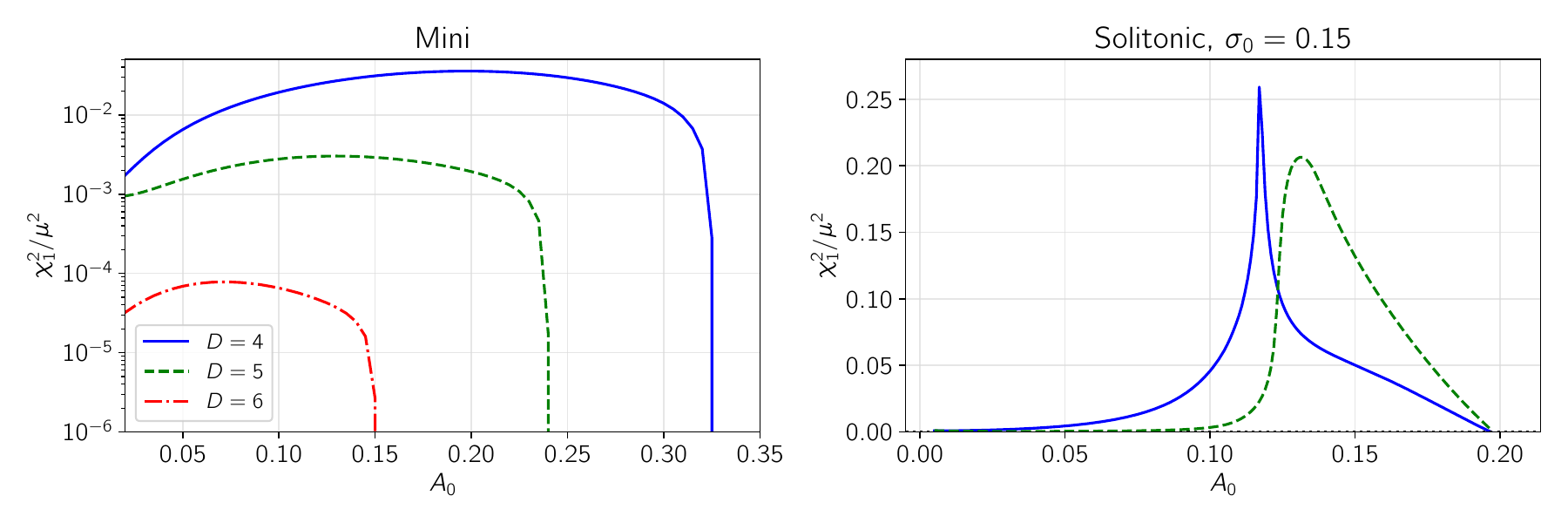}
 \caption{First excited radial frequency $\chi_1^2$ against central amplitude $A_0$, for mini and solitonic families.}
\label{fig:excited_modes}
\end{figure}

Finally, we note that our method is also capable of extracting overtone frequencies. As an example, in Fig.~\ref{fig:excited_modes} we plot the first overtone frequencies $\chi_1^2$ for a selection of families in the region in which they are positive. For mini BSs in $D = 5$ and $6$, the region in which $\chi_1^2 > 0$ corresponds to the branch between $A_0 = 0$ and the first minimum in the $M(A_0)$ curve. We thus conclude that mini boson stars in $D \in \{5, 6 \}$ dimensions have precisely one unstable mode on the first branch, extending the results of Ref.~\cite{Franzin_2024}.

\section{Dynamical Evolutions} \label{dynamical}
While we have demonstrated the existence of higher-dimensional boson stars that are stable under radial perturbations, this by no means closes the book on the stability analysis of these models. Even just considering spherical dynamics, it is conceivable that nonlinear effects could spoil the stability of linearly stable models, or indeed re-stabilize models that suffer an unstable mode. This is to say nothing of potential nonspherical instabilities. In this section we take the first step towards a full understanding of the nonlinear dynamics by performing a series of dynamical evolutions in spherical symmetry. 

Our approach uses the code {\sc
sphericalbsevolver} ({\sc
SBSE}) \cite{Marks_Spherical_BS_Evolver}, which is based on dimensional reductions of the BSSN and CCZ4 systems commonly used in numerical relativity. A full description of the evolution system used is given in Appendix~\ref{evolution_scheme}, while convergence results are shown in Appendix~\ref{convergence}.
Finally, in Appendix~\ref{bssn_vs_ccz4} we compare results obtained using the BSSN and CCZ4 systems, showing that BSSN generally displays long-lasting linear growth in the Hamiltonian constraint, while with CCZ4 both constraint norms level off at a finite value. This may suggest CCZ4 to be a superior choice for long-term stability studies, as the unbounded accumulation of unphysical constraint-violating modes in the BS interior could potentially lead one to faulty conclusions regarding dynamical stability.

\subsection{Unperturbed Evolutions}
\begin{figure}
 \centering
        \includegraphics[width=\linewidth]{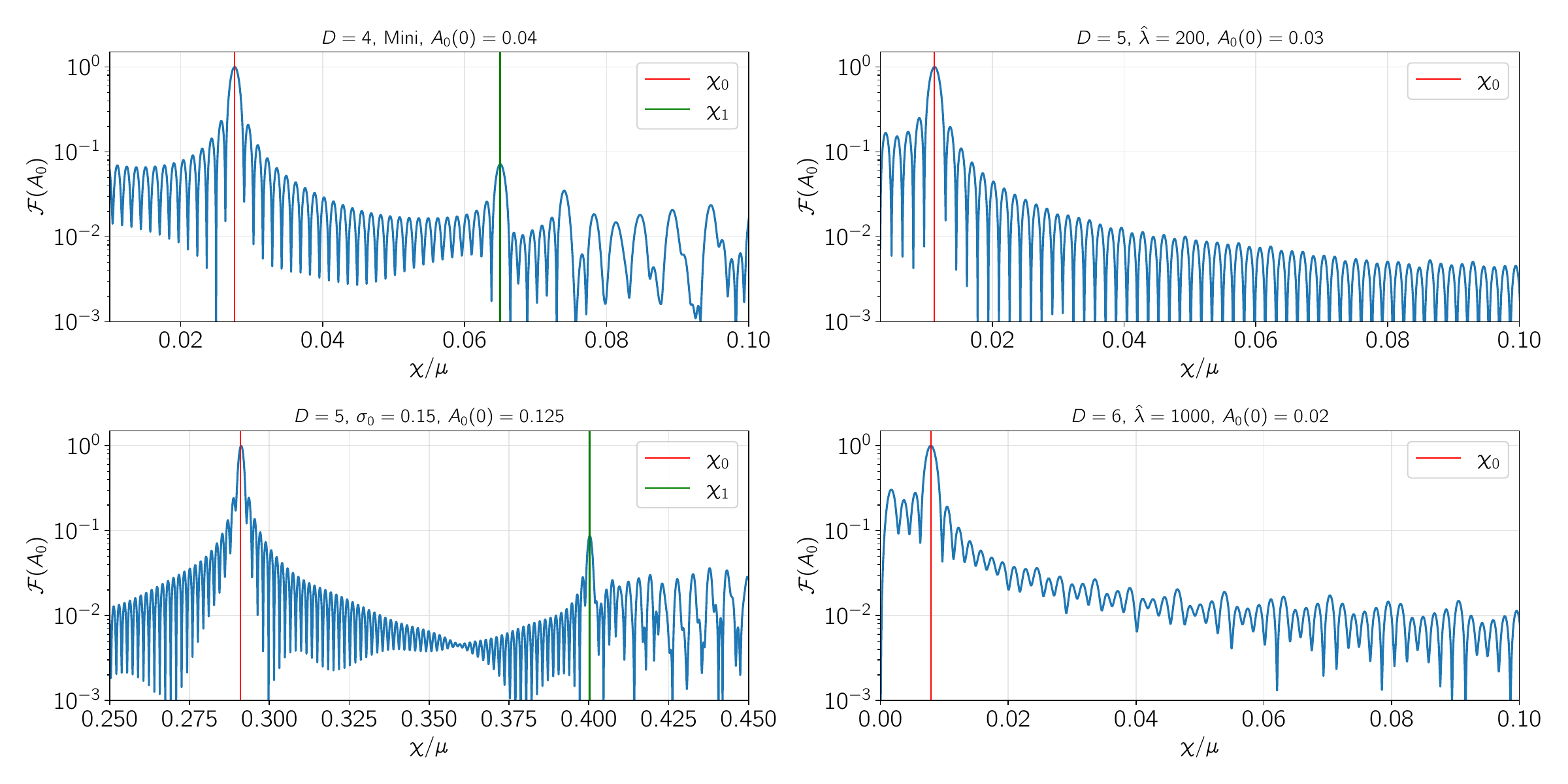}
 \caption{Power spectrum of the central amplitude $A_0$ over time for a selection of our unperturbed dynamical runs, compared to the fundamental radial mode $\chi_0$ and first overtone $\chi_1$ in cases where it is clearly excited. }
\label{fig:power_spectra}
\end{figure}

We begin by presenting a series of runs evolving BS models with only numerical perturbations due to truncation error. First, to independently test the accuracy of the radial oscillation frequencies computed in the previous section, we perform a set of longer runs for models found to be stable. We then compare the power spectrum of the central amplitude over time to the $\chi^2$ values determined by perturbative analysis. Results are shown in Fig.~\ref{fig:power_spectra}. We find very good agreement in all cases, providing evidence that both our computed oscillation frequencies and numerical method are robust. In general, local maxima in the power spectrum seen beyond the first overtone $\chi_1$ do not necessarily correspond to higher computed overtone frequencies; these may be due to physical oscillations having a nonlinear origin, or to noise propagating in from the outer boundary.

We then perform a larger set of evolutions along a selection of one-parameter families, corresponding to those families for which we displayed radial oscillation frequencies in Section~\ref{perturbative}. {Following Ref.~\cite{Marks:2025}}, where an instability is {found} we define the instability timescale as the time at which the central amplitude $A_0(t)$ departs by $10\,\%$ from its original value. Models run for a time of at least $t \sim 10^4 M^{1/(D - 3)}$ with no sign of instability are designated stable. We emphasize that the specific timescales obtained in this way will depend on a variety of factors such as resolution, gauge, evolution method and numerical parameter choices; within each family the specific $t_\mathrm{instability}$ values should therefore only be interpreted as having meaning relative to one another.

\begin{figure}[t]
 \centering
        \includegraphics[width=\linewidth]{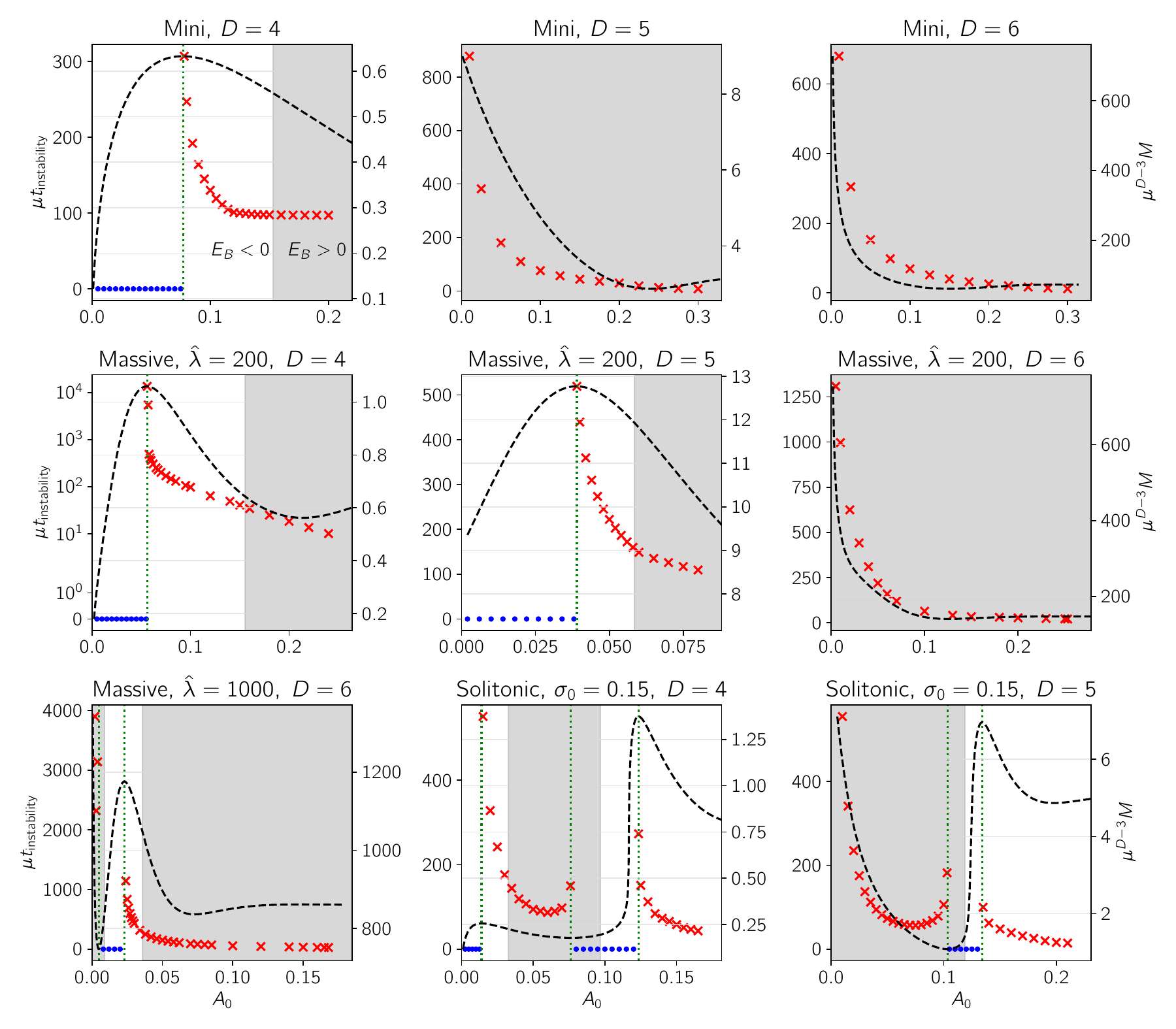}
 \caption{In each plot, the black dashed curve shows $M$ against $A_0$ for a BS family, corresponding to the scale on the right axis. The left axis shows the results of dynamical evolutions, using red crosses for unstable evolutions and showing the instability timescale, while stable evolutions use blue dots and are assigned the value $t_\mathrm{instability} = 0.$ Dotted vertical lines are placed at extrema of the $M(A_0)$ curve that correspond to a zero crossing of the fundamental radial oscillation frequency $\chi_0^2$, to emphasize that our instability timescales consistently show signs of diverging at these points. Shaded regions have positive binding energy.} 
\label{fig:instability_timescales}
\end{figure}

The results of these runs are summarized in Fig.~\ref{fig:instability_timescales}. There, we overlay the instability timescales obtained for each family with an $M(A_0)$ plot to emphasize the key point: in all cases the results of our dynamical evolutions agree with the predictions of linear theory, with instability timescales showing signs of diverging near the extrema of the $M(A_0)$ curve that correspond to zero crossings of $\chi_0^2$. Thus, as has been previously found in $D = 4$ \cite{Seidel_Suen_1990,Balakrishna_1998,Guzman_2009}, linear stability theory appears sufficient to predict the stability of higher-dimensional boson stars under nonlinear spherical dynamics. In addition, we shade regions of the parameter space for which the binding energy is positive. We draw particular attention to the presence of stable models with positive binding energy for solitonic families in $D = 4,5$ with $\sigma_0 = 0.15$, and for massive families in $D = 6$ with $\hat\lambda = 1000$. This phenomenon was noted previously in Ref.~\cite{Marks_2025_CP}, where it was found that even under explicit perturbations such models remained dynamically stable in spherical symmetry. Now having additional examples of this behavior in higher dimensions, it would be interesting to determine whether dispersive nonspherical instabilities appear when only an $SO(D - 2)$ or $SO(D - 3)$ symmetry is assumed. Where we do observe dynamical instabilities, we consistently find that, in accordance with previous work in $D = 4$~\cite{Guzman_2004,Guzman_2009,Alcubierre_2019}, the character of instability depends on the sign of the binding energy: models with $E_B < 0$ migrate back to the stable branch or collapse to BHs, while those with $E_B > 0$ disperse or collapse.

Finally, we take this opportunity to comment on a numerical matter we have observed in evolving higher-dimensional BS spacetimes. Our code uses the method of lines with a Runge-Kutta time integrator, with fourth-order discretization in both time and space. The Courant factor, defined as $\mathcal C:= \Delta t / \Delta x$ where $\Delta t$ and $\Delta x$ are the spatial and time step sizes respectively, must in general be smaller than $0.5$ to obtain numerically stable evolutions. In $D = 4$ and $5$, we consistently find that $\mathcal C = 0.4$ leads to such evolutions. However, we find that in $D = 6$, running with this value, particularly for strong self-interactions, can lead to numerical instabilities that do not disappear with increased resolution or grid size. In such cases we instead run with $\mathcal C = 0.25$; we have not found any models for which this choice is numerically problematic. The need to decrease the Courant factor when evolving higher-dimensional spacetimes has previously been observed in the case of black holes \cite{Cook_2017}.

\subsection{Perturbed Evolutions}
 We now present a series of runs in which we introduce explicit spherical perturbations to our BS models, 
\begin{equation}
    \varphi =\varphi_\mathrm{b} + \delta\varphi, \quad \Pi_\varphi = (\Pi_\varphi)_\mathrm{b} + \delta \Pi_\varphi
\end{equation}
where $\Pi_\varphi$ is the scalar field momentum, and the subscript $\mathrm{b}$ refers to the unperturbed background solution. Following the arguments of Ref.~\cite{Alcubierre_2019}, we consider perturbations that, to leading order in the perturbation size, conserve the Noether charge of the background solution. This can be accomplished by two methods:

\begin{enumerate}
    \item We take an \textit{internal} perturbation, contained within the BS interior, but conditioning the perturbation to the field momentum so that
    \begin{equation}
        \delta\Pi_\varphi = -\frac{i\omega}{\alpha_\mathrm{b}}\delta\varphi,
    \end{equation}
    where $\alpha_0$ is the initial lapse.
    \item We take an \textit{external} perturbation, which has compact support (effectively) outside the BS radius. This essentially models the infall of a spherical shell of scalar matter.
\end{enumerate}
In both cases, we prescribe perturbations using a Gaussian profile of the form $\delta \varphi = a \exp\left((r - r_0)^2 / k^2\right)$ for real constants $a, r_0, k$. Once we have perturbed our background BS, we also re-solve the Hamiltonian constraint {Eq.}~\eqref{eq:X_eq} and polar slicing condition {Eq.}~\eqref{eq:alpha_eq} to obtain new functions $X(r)$ and $\alpha(r)$ for use in our initial data.

Because there is already a robust literature concerning the spherical dynamics of BSs in $D=4,$ and we have found our numerical approach to agree qualitatively with the conclusions of this literature for unperturbed evolutions, here we focus specifically on the cases $D = 5$ and $D = 6$. Our results are summarized in Tables~\ref{tab:D5perts} and~\ref{tab:D6perts}. As with our unperturbed evolutions, we see that the predictions of linear stability theory are in all cases confirmed: models with $\chi_0^2 >0$ are dynamically stable under both types of perturbation, while those with $\chi_0^2 < 0$ always migrate, disperse, or collapse. 

\begin{table}
\caption{Summary of results for explicitly perturbed evolutions in $D = 5$. For each model we show the potential used (giving $\hat\lambda$ values for massive and $\sigma_0$ for solitonic families), central amplitude $A_0$, mass $M$, binding energy $E_B$, perturbation type (I for internal, II for external), perturbation parameters $a,k,r_0$, whether the model is on a perturbatively stable (S) or unstable (U) branch, and the dynamical outcome: S for stable evolution, D for dispersion, M for migration to a perturbatively stable branch, and BH for collapse to a black hole. }
\centering
\vspace{5pt}
\begin{tabular}{l c c c c c c c c c c}
\hline
Label & $V(|\varphi|^2)$ & $A_0$ & $\mu^2M$ & $\mu^2E_B$ & Type & $a$ & $k$ & $r_0$ & Branch & Result \\
\hline
\texttt{D5MAI} &Mini& 0.05 & 6.49 & 0.433 & I & -0.001 & 1.0 & 5.0 & U & D \\
\texttt{D5MAII} &Mini& 0.05 & 6.49 & 0.433 & II & 0.0005 & 1.0 & 25.0 & U & BH \\
\texttt{D5MBI} &Mini& 0.1 & 4.76 & 0.119 & I & -0.001 & 1.0 & 5.0 & U & BH \\
\texttt{D5MBI} &Mini& 0.1 & 4.76 & 0.119 & II & 0.0005 & 1.0 & 25.0 & U & BH \\
\texttt{D5LAI} &$\hat\lambda = 200$& 0.03 & 12.6 & -0.0460 & I & -0.001 & 1.0 & 3.0 & S & S \\
\texttt{D5LAII} &$\hat\lambda = 200$& 0.03 & 12.6 & -0.0460 & II & 0.0001 & 1.0 & 30.0 & S & S \\
\texttt{D5LBI} &$\hat\lambda = 200$& 0.05 & 12.5 & -0.0380 & I & -0.001 & 1.0 & 3.0 & U & M \\
\texttt{D5LBII} &$\hat\lambda = 200$& 0.05 & 12.5 & -0.0380 & II & 0.0005 & 1.0 & 30.0 & U & BH \\
\texttt{D5LCI} &$\hat\lambda = 1000$& 0.05 & 32.6 & 0.0538 & I & -0.001 & 1.0 & 3.0 & U & BH \\
\texttt{D5LCII} &$\hat\lambda = 1000$& 0.05 & 32.6 & 0.0538 & II & 0.0005 & 1.0 & 30.0 & U & BH \\
\texttt{D5SAI} & $\sigma_0 = 0.15$& 0.1 & 1.09 & 0.0685 & I & -0.001 & 2.0 & 5.0 & U & D \\
\texttt{D5SAII} & $\sigma_0 = 0.15$& 0.1 & 1.09 & 0.0685 & II & 0.001 & 1.0 & 30.0 & U & D \\
\texttt{D5SBI} & $\sigma_0 = 0.15$& 0.11 & 1.11 & 0.0661 & I & -0.001 & 1.0 & 3.0 & S & S \\
\texttt{D5SBII} & $\sigma_0 = 0.15$& 0.11 & 1.11 & 0.0661 & II & 0.0005 & 1.0 & 20.0 & S & S \\
\texttt{D5SCI} & $\sigma_0 = 0.15$& 0.14 & 6.74 & -3.06 & I & -0.001 & 2.0 & 0.0 & U & M \\
\texttt{D5SCII} & $\sigma_0 = 0.15$& 0.14 & 6.74 & -3.06 & II & 0.001 & 1.0 & 15.0 & U & BH \\
\texttt{D5SDI} & $\sigma_0 = 0.08$& 0.062 &145  & -353 & I & -0.002 & 2.0 & 5.0 & S & S \\
\texttt{D5SDII} & $\sigma_0 = 0.08$& 0.062 &145  & -353 & II & 0.001 & 1.0 & 30.0 & S & S \\
\hline
\end{tabular}
\label{tab:D5perts}
\end{table}

\begin{table}
\caption{ Summary of results for explicitly perturbed evolutions in $D = 6$. See caption of 
Table~\ref{tab:D5perts} for more details.}
\vspace{5pt}
\centering
\begin{tabular}{l c c c c c c c c c c}
\hline
Label & $V(|\varphi|^2)$ & $A_0$ & $\mu^3M$ & $\mu^3E_B$ & Type & $a$ & $k$ & $r_0$ & Branch & Result \\
\hline
\texttt{D6MAI} &Mini& 0.05 & 115 & 1.47 & I & -0.001 & 1.0 & 5.0 & U & D \\
\texttt{D6MAII} &Mini& 0.05 & 115 & 1.47 & II & 0.0005 & 1.0 & 30.0 & U & BH \\
\texttt{D6MBI} &Mini& 0.1 & 70.7 & 1.79 & I & -0.001 & 1.0 & 5.0 & U & BH \\
\texttt{D6MBI} &Mini& 0.1 & 70.7 & 1.79 & II & 0.0005 & 1.0 & 30.0 & U & BH \\
\texttt{D6LAI} &$\hat\lambda = 1000$& 0.006 & 756 & 0.421& I & 0.001 & 1.0 & 3.0 & S & S \\
\texttt{D6LAII} &$\hat\lambda = 1000$& 0.006 & 756 & 0.421 & II & 0.0001 & 0.5 & 60.0 & S & S \\
\texttt{D6LBI} &$\hat\lambda = 1000$& 0.02 & 1160 & -3.85& I & -0.001 & 1.0 & 5.0 & S & S \\
\texttt{D6LBII} &$\hat\lambda = 1000$& 0.02 & 1160 & -3.85& II & 0.0001 & 1.0 & 40.0 & S & S \\
\texttt{D6LCI} &$\hat\lambda = 1000$& 0.03 & 1123 & -2.71& I & -0.001 & 1.0 & 5.0 & U & M \\
\texttt{D6LCII} &$\hat\lambda = 1000$& 0.03 & 1123 & -2.71& II & 0.0001 & 1.0 & 40.0 & U & BH \\
\texttt{D6LDI} &$\hat\lambda = 1000$& 0.04 & 992 & 1.51& I & -0.001 & 1.0 & 5.0 & U & D \\
\texttt{D6LDII} &$\hat\lambda = 1000$& 0.04 & 992 & 1.51& II & 0.0001 & 1.0 & 40.0 & U & BH \\
\texttt{D6LEI} &$\hat\lambda = 200$&  0.05 & 208 & 2.00& I & -0.001 & 1.0 & 3.0 & U & D \\
\texttt{D6LEII} &$\hat\lambda = 200$& 0.05 & 208 & 2.00 & II & 0.0001 & 0.5 & 50.0 & U & BH \\
\texttt{D6SAI} &$\sigma_0 = 0.15$&  0.03 & 118 & 0.764 & I & 0.001 & 1.0 & 3.0 & U & D \\
\texttt{D6SAII} &$\sigma_0 = 0.15$& 0.03 & 118 & 0.764 & II & 0.0001 & 0.5 & 50.0 & U & D \\
\hline
\end{tabular}
\label{tab:D6perts}
\end{table}

The dynamical fate of unstable models is also determined in part by the sign of the binding energy: models with $E_B > 0$ never migrate; models with $E_B < 0$ never disperse. Whether an unstable model will collapse to a BH is harder to predict, however, as this can occur for both gravitationally bound and unbound solutions. There appears to be a dependency for some models on the type of perturbation: perturbations that add scalar matter are more likely to cause collapse to a BH, while those that remove it are less likely. For some models, though, both types of perturbation result in collapse; for others both types result in dispersion. A full understanding of this relationship requires further study. 

We also draw attention to the runs \texttt{D5SBI}, \texttt{D5SBII}, \texttt{D6LAI} and \texttt{D6LAII}, where models with positive binding energy are nonetheless seen to be dynamically stable under explicit perturbations both internal and external. This is true even for perturbations that further increase the binding energy relative to that of the equilibrium configuration. We therefore interpret these models as exhibiting a kind of metastability: stable to general spherical perturbations once formed, but unlikely to form generically under gravitational collapse of scalar clouds. The precise nature of this metastability, including robustness under nonspherical dynamics, requires further study.

\begin{figure}
 \centering
        \includegraphics[width=\linewidth]{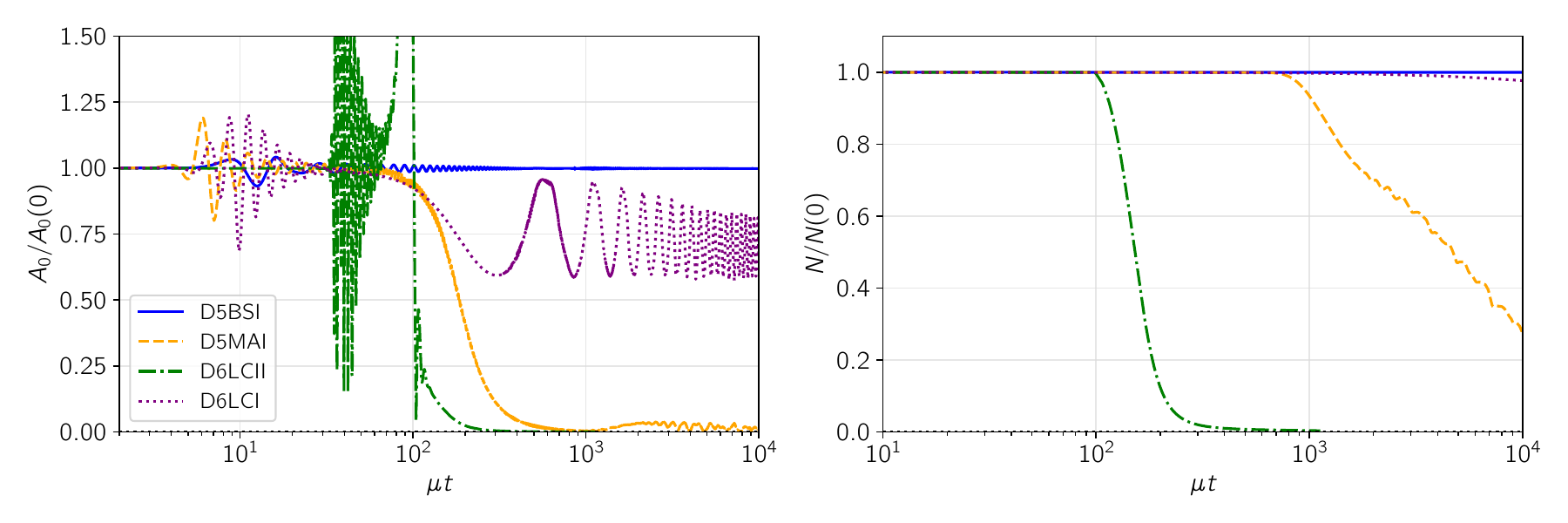}
 \caption{Central amplitude $A_0$ (left) and Noether charge $N$ (right) over time relative to their initial values for runs representative of our four dynamical fates: \texttt{D5SBI} (stable evolution), \texttt{D5MAI} (dispersion), \texttt{D6LCII} (collapse to black hole), and \texttt{D6LCI} (migration to stable branch).  }
\label{fig:evolution_examples}
\end{figure}

In Fig.~\ref{fig:evolution_examples}, we show the central amplitude and Noether charge relative to the equilibrium configuration for a selection of four runs corresponding to the four dynamical fates we observe: stable evolution, dispersion, migration to a stable branch, and collapse to a BH. The general features of each plot are generic for their respective dynamical fates. In particular, for migrating models, we see large, long-lasting oscillations that are damped on very long timescales relative to the oscillation frequency, with ejection of only a small fraction of the Noether charge. This is in contrast to dispersing models, for which the overall Noether charge in the computational domain exhibits a roughly exponential decay. In all cases, we see a small linear decay in the central amplitude and Noether charge over time. As we will show in Appendix~\ref{convergence}, however, this decay converges away as we increase the grid resolution, and we can therefore safely attribute it to mere truncation error.

\section{Conclusions}
\begin{figure}
 \centering
        \includegraphics[width=400pt]{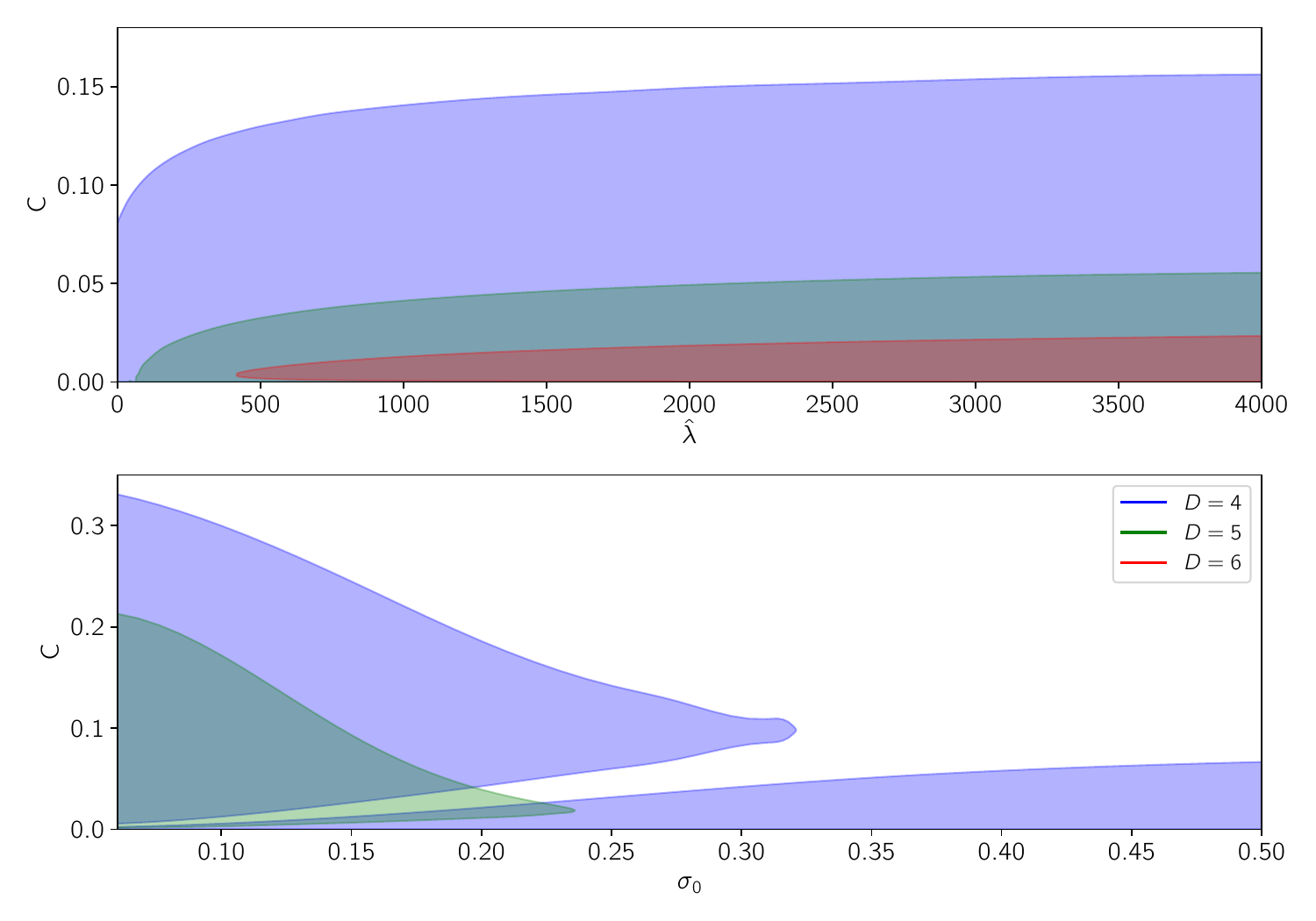}
 \caption{Boson star compactness, as defined in Eq.~\eqref{eq:C_def}, against the potential parameters $\hat\lambda$ and $\sigma_0$, where shaded regions include models we have identified as radially stable, for massive (top) and solitonic (bottom) families.   }
\label{fig:heatmap}
\end{figure}

We have now completed a detailed study of mini, massive and solitonic BS solutions in $D \in \{4,5,6\}$ spacetime dimensions, focusing especially on the radial stability of these solutions by means of both perturbative analysis and nonlinear evolution. Our main results can be summarized as follows:
\begin{itemize}
    \item A self-interacting (in $D = 5,6$) or solitonic (in $D = 5$) potential can be used to obtain radially stable higher-dimensional BS models. This is confirmed by both the linear and nonlinear analyses. 
    \item Families of BSs with solitonic potential in $D = 6$ display a pathological divergence at finite central amplitude for sufficiently small $\sigma_0$, and are in all cases radially unstable.
    \item The sign of the binding energy is neither necessary nor sufficient to predict the radial stability of BS models in $D \ge 4$ in general. While it can constrain the dynamical fates of models on an unstable branch, it cannot fully determine these.
\end{itemize}
We also note that the stable self-interacting models we have identified can reach arbitrarily low compactness, for any $\hat\lambda$ above the critical value in $D = 5$ and by choice of sufficiently large $\hat\lambda$ in $D = 6$. Given our intuition that the faster $1 / r^{D - 3}$ decay of the gravitational field in higher dimensions suggests a general instability to fission, one might consider this fact surprising. In Fig.~\ref{fig:heatmap}, we indicate the range of compactness values achievable by stable models for each BS family we consider in this work.

The most obvious avenue for future work is an analysis of the nonradial stability of models we have identified as stable in this paper. This would be especially interesting for those models which are stable with positive binding energy, to determine the character of a potential nonspherical fissive instability. While dynamical evolutions with no enforced symmetry are likely out of reach for the foreseeable future due to the computational cost of evolving full $(4 + 1)$-dimensional spacetimes and beyond, evolutions that enforce only $SO(D - 3)$ or $SO(D - 2)$ symmetries are feasible. Even if radially stable models suffer instabilities due to modes that do not respect these symmetries, such instabilities would likely not prove problematic for our proposed applications in studying e.g. critical phenomena or the hoop conjecture in higher dimensions, where models would likely be evolved on a 3D, 2D, or indeed even 1D grid, employing the corresponding symmetry assumption. {A study of larger explicit perturbations would also be interesting, especially in the context of critically tuning the perturbation size. We plan to address this in follow-up work.} An additional avenue to explore could be a perturbative analysis that takes into account nonradial modes, though we expect such a project would be highly involved. Early results in this direction using axisymmetric perturbations have recently been obtained in Ref.~\cite{Liang_2025} in the case $D = 4$.

Apart from this, it would be interesting to determine the stability properties of other exotic compact object models such as Proca stars \cite{Brito_2016} in higher dimensions, and to investigate the effect of replacing general relativity with a higher-curvature gravity model such as Gauss-Bonnet gravity, which becomes topologically trivial in $D = 4$. Finally, there is the question of to what extent our results generalize for arbitrary $D > 6$. A possible framework for answering this question may be found in the large $D$ limit of general relativity \cite{Emparan_2013}.

\ack{G.A.M. is grateful for helpful discussions with Seppe Staelens, Tamara Evstafyeva, Christopher J. Moore, and especially to his supervisor Ulrich Sperhake. A.A.Z. is grateful to the Cambridge UG/PG Research Café for connecting to G.A.M for this project.
Computations were done on
the CSD3 and Fawcett (Cambridge), Cosma (Durham), Niagara (Toronto), Narval (Montreal),
Stampede2 (TACC) and Expanse (SDSC) clusters. We acknowledge support by the NSF Grant Nos.~PHY-090003,~PHY-1626190
and PHY-2110594, DiRAC projects ACTP284 and ACTP238, STFC capital
Grants Nos.~ST/P002307/1,~ST/R002452/1 and ST/V005618/1,
STFC operations Grant No.~ST/R00689X/1.
}

\appendix
\section{Evolution Scheme} \label{evolution_scheme}
\newcommand {\cgamma} {\tilde{\gamma}}

In this appendix, we present the evolution system used by {\sc SBSE}. The code uses the modified cartoon approach to dimensional reduction \cite{Pretorius_2004, Cook:2016soy}, projecting all physical quantities onto a particular coordinate axis, which we take to be the $z$-axis. We then use tensor transformation laws to replace partial derivatives in the other directions. In the present case of an $SO(D - 1)$ symmetry, we will take the spatial coordinates on a given time slice to be
\begin{equation}
x^i = (z, w^1, ..., w^{D - 2}).
\end{equation}
In addition, early Latin indices $a,b,c,...$ will be understood to run over the $D - 2$ off-axis coordinates, so that we can also write the above as $(z, w^a)$. In cases in which any of the last $D - 2$ coordinates could be used interchangeably, we simply write $w$. We begin with the BSSN formulation of numerical relativity, which starts with the standard ADM decomposition into lapse $\alpha$, shift vector $\beta^i$, and spatial $3$-metric $\gamma_{ij}$. It then introduces as evolution variables:
\begin{itemize}
\item The conformal factor $\chi = \gamma^{\frac{1}{D-1}},$
\item The conformally rescaled metric $\cgamma_{ij} = \chi\gamma_{ij},$
\item The mean curvature $K = \gamma^{ij}K_{ij},$
\item The conformally rescaled traceless extrinsic curvature $\tilde{A}_{ij} = \chi\left(K_{ij} - \frac{1}{D - 1} K\gamma_{ij}\right)$, and lastly
\item The contracted conformal Christoffel symbols, $\tilde{\Gamma}^i = \frac{1}{2}\gamma^{il}\gamma^{jk}\left(\partial_j \cgamma_{lk} + \partial_k \cgamma_{lj} - \partial_l \cgamma_{jk} \right). $
\end{itemize}
Furthermore, we have the usual  energy density, current vector, and stress tensor by projecting onto space and time directions as follows:
\begin{equation}
\rho = T_{\mu\nu}n^\mu n^\nu, \quad j_i = -\perp^\mu_{\;i}T_{\mu\nu}n^\nu, \quad S_{ij} = \perp^\mu_{\;i}\perp^\nu_{\;j}T_{\mu\nu},
\end{equation}
where $n = (\partial_t - \beta) / \alpha$ is the timelike unit normal vector and $\perp^\mu_{\;\nu} = \delta^\mu_\nu + n^\mu n_\nu$ the spatial projection operator. We also evolve the scalar field momentum,
\begin{equation}
     \Pi_\varphi := -\frac{1}{2}\mathcal{L}_n\varphi = -\frac{1}{2\alpha} \left(\partial_t\varphi - \beta^k\partial_k\varphi \right). \label{Pidef}
\end{equation}

We are now ready to present the modified cartoon version of the BSSN and matter equations in $SO(D - 1)$ symmetry. Note that symmetry considerations imply that the $w$ components of all vectors and the $zw$ components of all $(0,2)$ tensors vanish; therefore we have a diagonal metric $\cgamma_{ij} = \mathrm{diag}(\cgamma_{zz}, ..., \cgamma_{ww}).$ Defining $n := D - 2$ for simplicity and using $\mathrm{TF}$ to denote trace-free parts, the evolution equations are as follows:
\begin{align}
    \partial_t\chi = \, & \beta_z\partial_z\chi + \frac{2}{D - 1}\chi\left[\alpha K - \partial_z\beta^z - n\frac{\beta^z}{z}\right], \label{chieq} \\
    \partial_t\cgamma_{zz} = \, & \beta^z\partial_z\cgamma_{zz} + 2\cgamma_{zz}\partial_z\beta^z - \frac{2}{D - 1}\left[\partial_z\beta^z + n\frac{\beta^z}{z}  \right] - 2\tilde{A}_{zz},\\
    \partial_t\cgamma_{ww} = \, &\beta^z\partial_z\cgamma_{ww} -   \frac{2}{D - 1}\cgamma_{ww}\left[\partial_z\beta^z - \frac{\beta^z}{z}  \right] - 2\tilde{A}_{ww},\\
    \partial_t\tilde{A}_{zz} = \, & \beta^z\partial_z\tilde{A}_{zz} + 2\tilde{A}_{zz}\partial_z\beta^z - \frac{2}{D - 1}\tilde{A}_{zz}\left[\partial_z\beta^z + n\frac{\beta^z}{z} \right] + \alpha K \tilde{A}_{zz} - 2\alpha\tilde{A}_{zz} \tilde{A}^z_{\;z} \\
    &+ \chi\left[\mathcal{R}_{zz} - D_zD_z\alpha - 8\pi\alpha S_{zz} \right]^{\mathrm{TF}}, \nonumber\\
    \partial_t\tilde{A}_{ww} = \, & \beta^z\partial_z\tilde{A}_{ww} - \frac{2}{D - 1}\tilde{A}_{ww}\left[\partial_z\beta^z - \frac{\beta^z}{z} \right] + \alpha K \tilde{A}_{ww} - 2\alpha\tilde{A}_{ww} \tilde{A}^w_{\;w} \\
    &+ \chi\left[\mathcal{R}_{ww} - D_wD_w\alpha - 8\pi\alpha S_{ww} \right]^{\mathrm{TF}}, \nonumber\\
    \partial_tK = \, & \beta^z\partial_zK - \chi\cgamma^{zz}D_zD_z\alpha + \alpha\left(\tilde{A}^{zz}\tilde{A}_{zz} + n\tilde{A}^{ww}\tilde{A}_{ww}\right) + \frac{\alpha K^2}{D-1}+ n\chi\cgamma^{ww}D_wD_w\alpha \\
    \nonumber & + \frac{8\pi\alpha}{n} \left(S + (D - 3)\rho \right), \\
    \partial_t \tilde{\Gamma}^z = \, & \beta^z\partial_z\tilde{\Gamma}^z  + \frac{2}{D - 1}\tilde{\Gamma}^z\left[\partial_z\beta^z + n\frac{\beta^z}{z}  \right] + \cgamma^{zz}\partial_z\partial_z\beta^z + n\cgamma^{ww}\left[\frac{\partial_z\beta}{z} - \frac{\beta^z}{z^2}\right] \\ \nonumber
    & - \tilde{\Gamma}^z\partial_z\beta^z + \frac{D - 3}{D - 1} \left[\cgamma^{zz}\partial_z\partial_z\beta^z + n \cgamma^{zz}\left(\frac{\partial_z\beta^z}{z} - \frac{\beta^z}{z^2}\right)\right] - \frac{2n\alpha}{D - 1}\cgamma^{zz}\partial_zK \\ \nonumber &- \tilde{A}^{zz}\left[(D - 1)\alpha\frac{\partial_z\chi}{\chi} + 2\partial_z\alpha \right] + 2\alpha\left[\tilde{\Gamma}^z_{zz}\tilde{A}^{zz} + \tilde{\Gamma}^z_{ww}\tilde{A}^{ww} \right] - \sigma \partial_z\beta^z \mathcal{G}^z - 16\pi\alpha\cgamma^{zz}j_z, \\
     \partial_t\varphi = \, & -2\alpha \Pi_\varphi + \beta^z\partial_z\varphi, \\
     \partial_t\Pi_\varphi = \, & \beta^z\partial_z\Pi_\varphi + \alpha K \Pi_\varphi + \frac{1}{2}\alpha \frac{\du V}{\du |\varphi|^2}(|\varphi|^2)\varphi + \frac{D - 3}{4}\alpha \cgamma^{zz}\partial_z\chi\partial_z\varphi \label{Pphieq} \\ \nonumber
     &- \frac{1}{2}\chi \left[\cgamma^{zz}\partial_z\varphi\partial_z\alpha + \alpha\left(\cgamma^{zz} \tilde{D}_z\tilde{D}_z\varphi  + n\cgamma^{ww} \tilde{D}_w\tilde{D}_w\varphi \right) \right],
\end{align}
where $\sigma > 0$ is a damping parameter. We also need some additional quantities: first, the conformal Christoffel symbols, given as
\begin{align}
\tilde{\Gamma}^z_{zz} &= \frac{1}{2}\cgamma^{zz}\partial_z{\cgamma_{zz}}, \quad \tilde{\Gamma}^z_{wz} = 0,  \quad \tilde{\Gamma}^z_{ab} = \delta_{ab}\left[ \frac{1 - \cgamma^{zz}\cgamma_{ww}}{z} - \frac{1}{2}\cgamma^{zz}\partial_z\cgamma_{ww}\right], \\ \nonumber
\tilde{\Gamma}^w_{zz} &= 0, \quad \tilde{\Gamma}^a_{bz} = \frac{1}{2}\delta^a_\mathrm{b} \partial_z\cgamma_{ww}, \quad \tilde{\Gamma}^a_{bc} = 0.
\end{align}
One can similarly compute the conformal covariant derivatives, Ricci tensor components, and matter quantities straightforwardly, starting from the standard decomposition in terms of BSSN variables. The Hamiltonian and momentum constraint violations are given by the following expressions:
\begin{align}
\mathcal{H} = \, & \chi\cgamma^{zz}\mathcal{R}_{zz} + n\chi\cgamma^{ww}\mathcal{R}_{ww} + \frac{D - 2}{D - 1}K^2 + \tilde{A}^{zz}\tilde{A}_{zz} + n \tilde{A}^{ww}\tilde{A}_{ww} - 16\pi\rho \label{reduced_Ham}\\
\mathcal{M}_z = \, & -\frac{D - 2}{D - 1}\partial_zK + n \cgamma^{ww}\left[\frac{\tilde{A}_{zz} - \tilde{A}_{ww}}{z} - \frac12 \cgamma^{ww}\tilde{A}_{ww}\partial_z\cgamma_{ww} - \tilde{\Gamma}^z_{ww}\tilde{A}_{zz}\right] \\ \nonumber &+ \cgamma^{zz}\left[\partial_z \tilde{A}_{zz} - 2 \tilde{\Gamma}^z_{zz}\tilde{A}_{zz} - \frac{D - 1}{2\chi} \tilde{A}_{zz}\partial_z\chi\right] - 8\pi j_z.
\end{align}
While our evolution equations are lengthy, an advantage of this formalism is that, besides allowing us to evolve spacetimes in an arbitrary number of dimensions using the same infrastructure, we retain the full gauge freedom afforded by the $3+1$ decomposition.  For instance, while we use by default the standard $1 + \log$ slicing condition \cite{Bona_1995}, given in our case by
\begin{equation}
    \partial_t\alpha = \beta^z\partial_z\alpha - 2\alpha K,
\end{equation}
we have found that the shock-avoiding gauge condition \cite{Alcubierre_1997}, given by
\begin{equation}
    \partial_t\alpha = \beta^z\partial_z\alpha - (1 + \alpha^2) K,
\end{equation}
can improve constraint violations. For the shift, we can simply set $\beta^z = 0$ to obtain a simpler evolution scheme, but we generally see improvements when we instead use the integrated gamma-driver condition, given by 
\begin{equation}
    \partial_t \beta^z = \beta^z\partial_z\beta^z + \frac34 \tilde{\Gamma}^z - \eta \beta^z,
\end{equation}
finding that this tends to minimize overall constraint violations for $\eta \sim 2 / M^{\frac{1}{D - 3}}$ where $M$ is the boson-star mass. In addition, since we evolve the same quantities as in the standard BSSN equations, a quantitative comparison to the results of dynamical evolutions obtained with codes such as GRChombo \cite{Radia:2021smk} is straightforward. We expect this to be useful in future studies involving dynamical evolutions beyond spherical symmetry.

We now turn to the CCZ4 system, which is derived by appending to the action \eqref{eq:action} an additional term involving a vector field $Z^\mu$,
\begin{equation}
    S_Z = \frac{1}{16\pi}\int\du^{D}x \nabla_\mu Z^\mu,
\end{equation}
before performing a decomposition similar to that used in BSSN. The resulting system of equations is known to have constraint-damping properties \cite{Alic:2011gg, Alic:2013xsa}. Applying the modified cartoon method to the CCZ4 system, we end up with an evolution system very similar to that presented for BSSN~\eqref{chieq}--\eqref{Pphieq}. Introducing $\Theta = -n_\mu Z^\mu$ and $\Theta^i = Z^i + \frac{\Theta}{\alpha}\beta^i$, the equations for $\chi$ and $\cgamma_{ij}$ are as before, while the equations for $\tilde{A}_{ij}$ are simply modified by the addition of a term
\begin{equation}
    \partial_t\tilde{A}_{ij}\rightarrow \partial_t\tilde{A}_{ij} -2\alpha\Theta\tilde{A}_{ij}.
\end{equation}
The evolution equation for $K$ becomes
\begin{align}
    \partial_t K= \, & \beta \partial_z K - \chi \cgamma^{zz}D_zD_z\alpha - n\cgamma_{ww}\chi D_{w}D_w\alpha + \alpha \bigg[\chi\cgamma^{zz}\mathcal{R}_{zz} + n\chi\cgamma^{ww}\mathcal{R}_{ww}  
    \\ \nonumber&+ K(K - 2\Theta) - (D - 1)\kappa_1(1 + \kappa_2)\Theta + \frac{8\pi}{D - 2}\left(S - (D - 1)\rho\right)
    \bigg],
\end{align}
where $\kappa_1$ and $\kappa_2$ are damping parameters; we typically set $\kappa_2 = 0$ and use $\kappa_1 \sim 0.1 / M^{\frac{1}{D - 3}}$. Instead of evolving $\tilde{\Gamma}^z$ directly, we define $\hat\Gamma^z := \tilde{\Gamma}^z + \frac{2}{\chi}\Theta^z$ and use this as an evolution variable. It is evolved according to
\begin{align}
    \partial_t \hat{\Gamma}^z= \, & \beta^z\partial_z\hat{\Gamma}^z  + \frac{2}{D - 1}\tilde{\Gamma}^z\left[\partial_z\beta^z + n\frac{\beta^z}{z}  \right] + \cgamma^{zz}\partial_z\partial_z\beta^z + n\cgamma^{ww}\left[\frac{\partial_z\beta}{z} - \frac{\beta^z}{z^2}\right] \\ \nonumber
    & - \hat{\Gamma}^z\partial_z\beta^z + \frac{D - 3}{D - 1} \left[\cgamma^{zz}\partial_z\partial_z\beta^z + n \cgamma^{zz}\left(\frac{\partial_z\beta^z}{z} - \frac{\beta^z}{z^2}\right)\right] - \frac{2n\alpha}{D - 1}\cgamma^{zz}\partial_zK \\ \nonumber &- \tilde{A}^{zz}\left[(D - 1)\alpha\frac{\partial_z\chi}{\chi} + 2\partial_z\alpha \right] + 2\alpha\left[\tilde{\Gamma}^z_{zz}\tilde{A}^{zz} + \tilde{\Gamma}^z_{ww}\tilde{A}^{ww} \right] - 16\pi\alpha\cgamma^{zz}j_z \\ \nonumber
    & - \frac{4}{D - 1}\frac{\alpha}{\chi}K\Theta^z + 2\Theta\cgamma^{zz}\partial_z\alpha - \frac{2\kappa_1\alpha}{\chi}\Theta^z + \frac{\sigma}\chi \left(\frac{2n}{z}\beta^z - (D - 3)\partial_z\beta \right).
\end{align}
The matter evolution equations are unchanged. Finally, the evolution equation for $\Theta$ is given by
\begin{align}
    \partial_t \Theta = \, & \beta \partial_z \Theta + \frac{\alpha}{2}\bigg[ \chi \cgamma^{zz}\mathcal{R}_{zz} - \tilde{A}_{zz}\tilde{A}^{zz} + \frac{n}{D - 1}K^2 + n\left(\chi \cgamma^{zz}\mathcal{R}_{zz} - \tilde{A}_{ww}\tilde{A}^{ww} \right)
    \\ \nonumber& -2K\Theta - 2\Theta^z \frac{\partial_z\alpha}{\alpha} - \kappa_1(D + n\kappa_2)\Theta - 16\pi\rho
    \bigg].
\end{align}
{We also note that we use simple outgoing Sommerfeld boundary conditions at the outer boundary for all numerical fields,
\begin{equation}
    \partial_t \phi + \partial_r \phi + \frac{D - 2}{2}(\phi - \phi_\infty) = 0
\end{equation}
where $\phi$ can represent any grid variable with asymptotic value $\phi_\infty$.
}

\section{Convergence} \label{convergence}
In this appendix, we discuss the convergence of our dynamical evolutions. In general, we observe  convergence between third and fourth order of all physical quantities, settling down to approximately third-order at late times. This arises from the fourth-order discretizations scheme we apply in both space and time, with up to one order of convergence lost due to factors of $\frac{1}{z}$ present in our evolution equations which must be handled by use of a regularization scheme at the origin.\footnote{We verify this by running the code with these terms disabled, and checking that we then obtain the expected fourth-order convergence.}

\begin{figure}
 \centering
        \includegraphics[width=\linewidth]{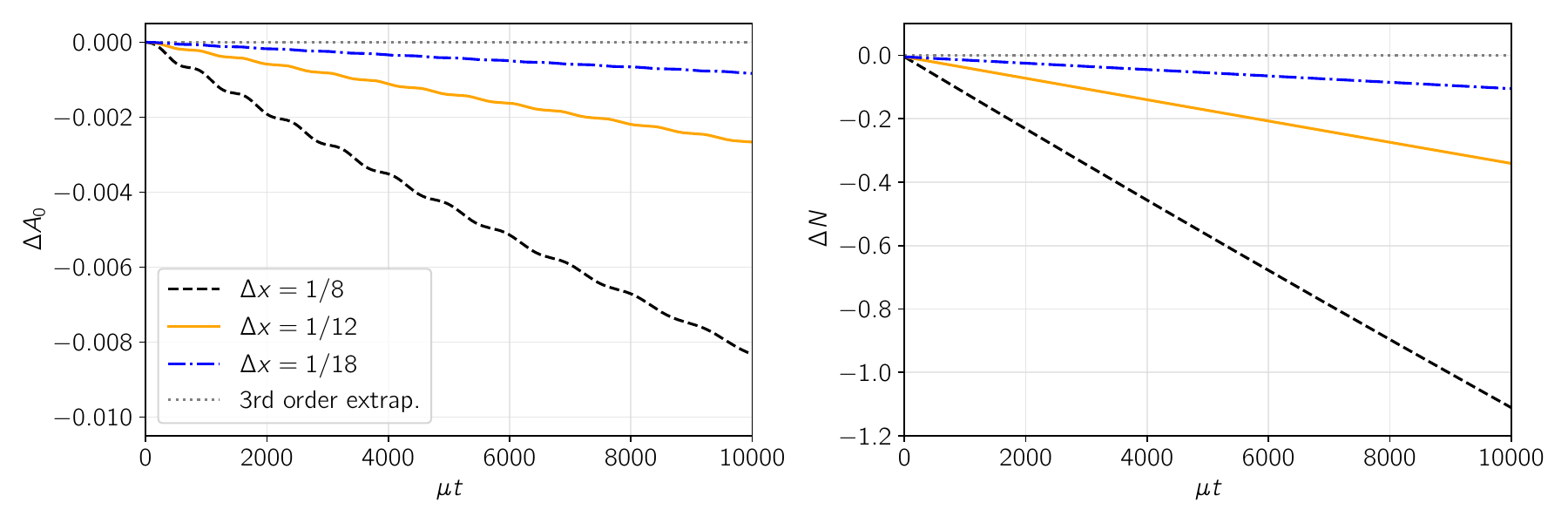}
 \caption{Relative decay in the central amplitude (left) and Noether charge (right) over time for a stable model with $\hat\lambda = 200$, $A_0 = 0.02$ in $D = 5$ dimensions at three resolutions, as well as the result of a third-order Richardson extrapolation showing convergence to zero decay.}
\label{fig:matter_convergence}
\end{figure}

First, in Fig.~\ref{fig:matter_convergence} we show the third-order convergence in the central amplitude and Noether charge for one of our stable models, with $\hat\lambda = 200$, $A_0 = 0.02$ in $D = 5$ dimensions. The linear decay in both quantities effectively vanishes under a third-order Richardson extrapolation, and so we conclude it is due simply to truncation error. This behaviour is generic for stable models.

\begin{figure} 
 \centering
        \includegraphics[width=\linewidth]{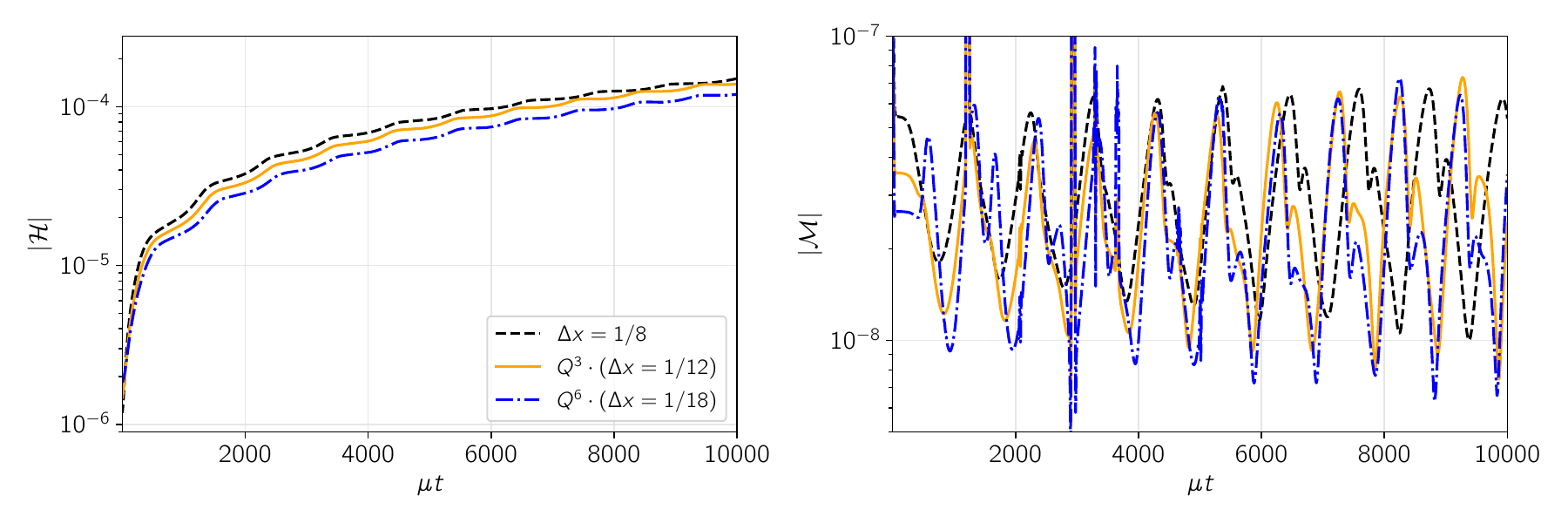}
 \caption{ $L_2$ norms of the Hamiltonian (left) and momentum (right) constraints over time  for a migrating model with $\hat\lambda = 200$, $A_0 = 0.05$ in $D = 5$ dimensions at three resolutions, with appropriate factors of the convergence factor $Q = 1.5$ to demonstrate third- to fourth-order convergence to zero constraint violation.    }
\label{fig:constraint_convergence}
\end{figure}

In addition, in Fig.~\ref{fig:constraint_convergence} we show an example of our constraint convergence for an unstable run, using a massive model with $\hat\lambda = 200$, $A_0 = 0.05$ in $D = 5$ dimensions using the BSSN formalism. The outcome in this case is migration. We see that the Hamiltonian constraint violation converges cleanly at between third and fourth order to zero. The momentum constraint profile varies more across resolutions, as the varying resolutions produce different numerical perturbations and hence different patterns of oscillation, but again we see that overall there is approximately third-order convergence, and indeed that the violation is several orders of magnitude subdominant to that of the Hamiltonian constraint. This behavior is generic, assuming the outer boundary is taken sufficiently large that error propagating inwards does not dominate (reflection of this propagating error at the innermost gridpoint is responsible for the spikes seen in the momentum constraint violation).

\begin{figure}[t]
 \centering
        \includegraphics[width=\linewidth]{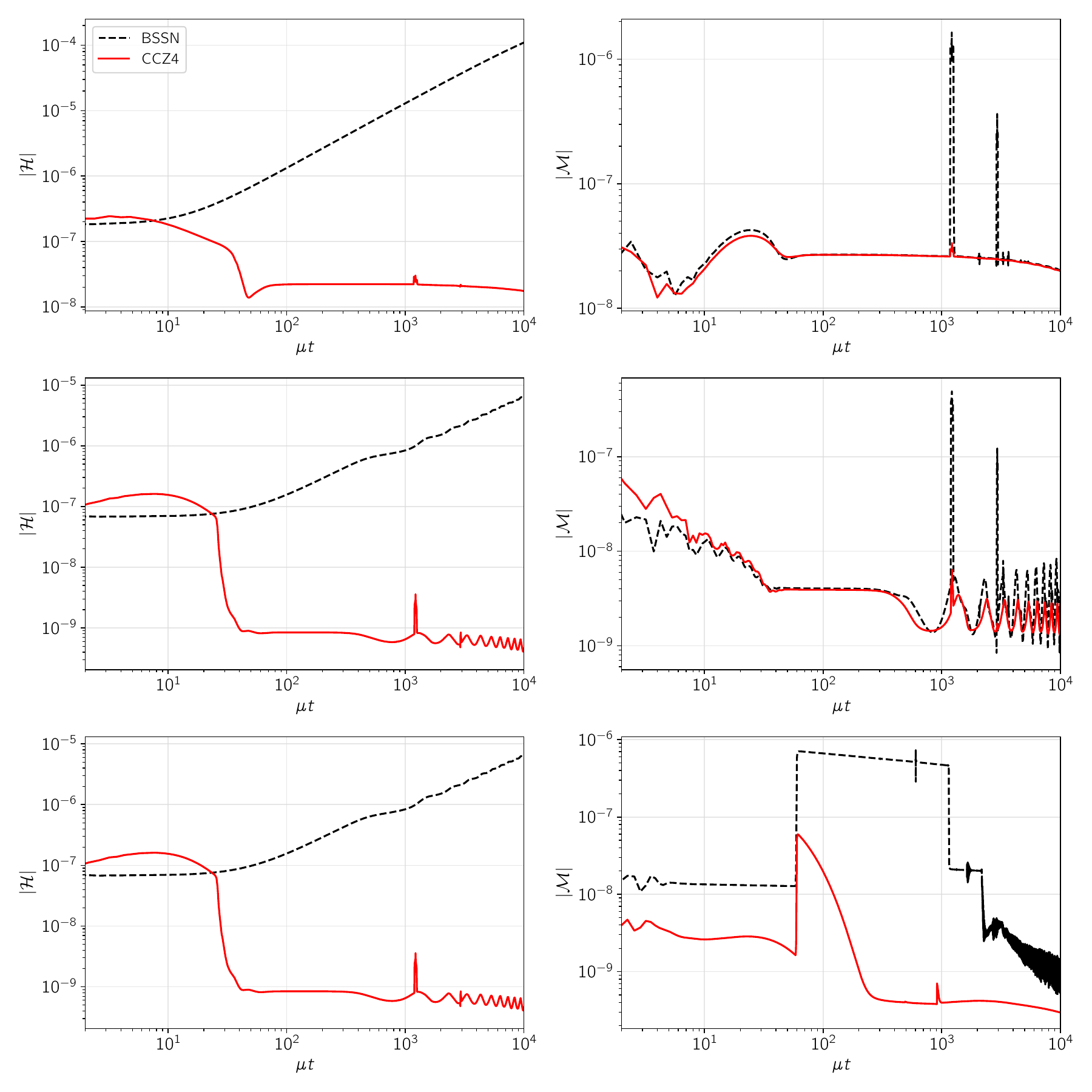}
 \caption{ $L_2$ norms of the Hamiltonian (left) and momentum (right) constraints over time for (top) a stable massive model with $A_0 = 0.02,$ $\hat\lambda = 200,$ $D = 6$, (middle) a migrating massive model with $A_0 = 0.05$, $\hat\lambda = 200$, $D = 5$, and (bottom) a stable solitonic model with $A_0 = 0.1$, $\sigma_0 = 0.15$, $D = 5$, showing the impact of switching between evolution using the BSSN and CCZ4 formalisms (with appropriately chosen damping parameters) with all other numerical parameters held constant   }
\label{fig:bssn_vs_ccz4}
\end{figure}

\section{Comparing BSSN to CCZ4}\label{bssn_vs_ccz4}
We see in Fig.~\ref{fig:constraint_convergence} that the Hamiltonian constraint displays long-lasting, approximately linear growth over the entire evolution. This occurs generically for both stable and migrating models (i.e. whenever scalar matter remains in the computational domain indefinitely) when using the BSSN formalism, regardless of damping factor,  Kreiss-Oliger dissipation, Courant factor, gauge conditions, or other numerical choices. In Fig.~\ref{fig:bssn_vs_ccz4}, we display some straightforward comparisons of the Hamiltonian and momentum constraints over time for BSSN and CCZ4 with appropriate damping parameters, showing that in CCZ4 both constraints settle down to finite values. Similar behaviour has been observed previously in the neutron-star case in four spacetime dimensions \cite{Alic:2011gg}. 

\funding{G.A.M. is supported by the Cambridge Trust at the University of Cambridge. A.A.Z has been supported by a Research and Professional Experiences Grant from Sidney Sussex College.}

\newpage
\bibliographystyle{iopart-num}
\bibliography{ref}

@article{Alic:2011gg,
  author = "Alic, D. and Bona-Casas, C. and Bona, C. and Rezzolla, L. and Palenzuela, C.",
  title = "{Conformal and covariant formulation of the Z4 system with constraint-violation damping}",
  journal = "Phys. Rev. D",
  volume = "85",
  pages = "064040",
  doi = "10.1103/PhysRevD.85.064040",
  year = "2012",
  note = "arXiv:1106.2254 [gr-qc]"}

@article{Alic:2013xsa,
  author = "Alic, D. and Kastaun, W. and Rezzolla, L.",
  title = "{Constraint damping of the conformal and covariant formulation of the Z4 system in simulations of binary neutron stars}",
  journal = "Phys. Rev. D",
  volume = "88",
  year = "2013",
  number = "6",
  pages = "064049",
  doi = "10.1103/PhysRevD.88.064049",
  eprint = "arXiv:1307.7391 [gr-qc]",
  reportNumber = "AEI-2013-231"}

@article{Baumgarte:1998te,
  author = "Baumgarte, T. W. and Shapiro, S. L.",
  title = "On the {N}umerical integration of {E}instein's field equations",
  journal = "Phys. Rev. D",
  year = "1998",
  doi = "10.1103/PhysRevD.59.024007",
  volume = "59",
  pages = "024007",
  note = "gr-qc/9810065"}

@article{Cardoso:2008bp,
   author = "Cardoso, V. and Miranda, A. S. and Berti, E. and Witek, H. and Zanchin, V. T.",
   title = "{Geodesic stability, Lyapunov exponents and quasinormal modes}",
   journal = "Phys. Rev. D",
   volume = "79",
   pages = "064016",
   doi = "10.1103/PhysRevD.79.064016",
   year = "2009",
   note = "arXiv:0812.1806 [hep-th]"}

@article{Collodel:2022jly,
  author = "Collodel, L. G. and Doneva, D. D.",
  title = "{Solitonic boson stars: Numerical solutions beyond the thin-wall approximation}",
  eprint = "2203.08203",
  archivePrefix = "arXiv",
  primaryClass = "gr-qc",
  doi = "10.1103/PhysRevD.106.084057",
  journal = "Phys. Rev. D",
  volume = "106",
  number = "8",
  pages = "084057",
  year = "2022"}

@article{Cook:2016soy,
  author = "Cook, William G. and Figueras, Pau and Kunesch, Markus and Sperhake, Ulrich and Tunyasuvunakool, Saran",
  title = "{Dimensional reduction in numerical relativity: Modified cartoon formalism and regularization}",
  journal = "Int. J. Mod. Phys. D",
  volume = "25",
  year = "2016",
  pages = "1641013",
  doi = "10.1142/S0218271816410133",
  note = "arXiv:1603.00362 [gr-qc]"}

@article{Cunha:2017qtt,
  author = "Cunha, P. V. P. and Berti, E. and Herdeiro, C. A. R.",
  title = "{Light-Ring Stability for Ultracompact Objects}",
  journal = "Phys. Rev. Lett.",
  volume = "119",
  year = "2017",
  number = "25",
  pages = "251102",
  doi = "10.1103/PhysRevLett.119.251102",
  eprint = "arXiv:1708.04211 [gr-qc]"}

@article{Cunha:2020azh,
  author = "Cunha, P. V. P. and Herdeiro, C. A. R.",
  title = "{Stationary black holes and light rings}",
  eprint = "2003.06445",
  archivePrefix = "arXiv",
  primaryClass = "gr-qc",
  doi = "10.1103/PhysRevLett.124.181101",
  journal = "Phys. Rev. Lett.",
  volume = "124",
  number = "18",
  pages = "181101",
  year = "2020"}

@article{Cunha:2022gde,
  author = "Cunha, P. V. P. and Herdeiro, C. and Radu, E. and Sanchis-Gual, N.",
  title = "{Exotic Compact Objects and the Fate of the Light-Ring Instability}",
  eprint = "2207.13713",
  archivePrefix = "arXiv",
  primaryClass = "gr-qc",
  doi = "10.1103/PhysRevLett.130.061401",
  journal = "Phys. Rev. Lett.",
  volume = "130",
  number = "6",
  pages = "061401",
  year = "2023"}

@article{Evstafyeva:2023kfg,
  author = "Evstafyeva, T. and Rosca-Mead, R. and Sperhake, U. and Bruegmann, B.",
  title = "{Boson stars in massless and massive scalar-tensor gravity}",
  eprint = "2310.05200",
  archivePrefix = "arXiv",
  primaryClass = "gr-qc",
  doi = "10.1103/PhysRevD.108.104064",
  journal = "Phys. Rev. D",
  volume = "108",
  number = "10",
  pages = "104064",
  year = "2023"}

@article{Evstafyeva:2024qvp,
  author = "Evstafyeva, T. and Sperhake, U. and Romero-Shaw, I. and Agathos, M.",
  title = "{Gravitational-wave data analysis with high-precision numerical relativity simulations of boson star mergers}",
  eprint = "2406.02715",
  archivePrefix = "arXiv",
  primaryClass = "gr-qc",
  doi = "10.1103/PhysRevLett.133.131401",
  journal = "Phys. Rev. Lett.",
  volume = "133",
  number = "13",
  pages = "131401",
  note = "arXiv:2406.02715 [gr-qc]",
  year = "2024"}

@article{Gleiser:1988ih,
  author = "Gleiser, Marcelo and Watkins, Richard",
  title = "{Gravitational Stability of Scalar Matter}",
  reportNumber = "NSF-ITP-88-152",
  doi = "10.1016/0550-3213(89)90627-5",
  journal = "Nucl. Phys. B",
  volume = "319",
  pages = "733--746",
  year = "1989"}

@article{Kain:2021rmk,
  author = "Kain, B.",
  title = "{Boson stars and their radial oscillations}",
  eprint = "2106.01740",
  archivePrefix = "arXiv",
  primaryClass = "gr-qc",
  doi = "10.1103/PhysRevD.103.123003",
  journal = "Phys. Rev. D",
  volume = "103",
  number = "12",
  pages = "123003",
  year = "2021"}

@article{Keir:2014oka,
  author = "Keir, J.",
  title = "{Slowly decaying waves on spherically symmetric spacetimes and ultracompact neutron stars}",
  eprint = "1404.7036",
  archivePrefix = "arXiv",
  primaryClass = "gr-qc",
  doi = "10.1088/0264-9381/33/13/135009",
  journal = "Class. Quant. Grav.",
  volume = "33",
  number = "13",
  pages = "135009",
  year = "2016"}

@article{Koga:2022dsu,
  author = "Koga, Y. and Asaka, N. and Kimura, M. and Okabayashi, K.",
  title = "{Dynamical photon sphere and time evolving shadow around black holes with temporal accretion}",
  eprint = "2202.00201",
  archivePrefix = "arXiv",
  primaryClass = "gr-qc",
  reportNumber = "OCU-PHYS-556, AP-GR-177, RUP-22-2",
  doi = "10.1103/PhysRevD.105.104040",
  journal = "Phys. Rev. D",
  volume = "105",
  number = "10",
  pages = "104040",
  year = "2022"}

@article{Radia:2021smk,
  author = "Radia, M. and Sperhake, U. and Drew, A. and Clough, K. and Figueras, P. and Lim, E. A. and Ripley, J. L. and Aurrekoetxea, J. C. and Fran\c{c}a, T. and Helfer, T.",
  title = "{Lessons for adaptive mesh refinement in numerical relativity}",
  eprint = "2112.10567",
  archivePrefix = "arXiv",
  primaryClass = "gr-qc",
  reportNumber = "KCL-PH-TH/2021-89",
  doi = "10.1088/1361-6382/ac6fa9",
  journal = "Class. Quant. Grav.",
  volume = "39",
  number = "13",
  pages = "135006",
  year = "2022"}

@article{Shibata:1995we,
  author = "Shibata, M. and Nakamura, T.",
  title = "Evolution of three-dimensional gravitational waves: {H}armonic slicing case",
  journal = "Phys. Rev. D",
  year = "1995",
  doi = "10.1103/PhysRevD.52.5428",
  volume = "52",
  pages = "5428-5444"}

@article{Synge:1966okc,
  author = "Synge, J. L.",
  title = "{The Escape of Photons from Gravitationally Intense Stars}",
  doi = "10.1093/mnras/131.3.463",
  journal = "Mon. Not. Roy. Astron. Soc.",
  volume = "131",
  number = "3",
  pages = "463--466",
  year = "1966"}

@article{Volkel:2022khh,
  author = {V\"olkel, S. H. and Franchini, N. and Barausse, E. and Berti, E.},
  title = "{Constraining modifications of black hole perturbation potentials near the light ring with quasinormal modes}",
  eprint = "2209.10564",
  archivePrefix = "arXiv",
  primaryClass = "gr-qc",
  doi = "10.1103/PhysRevD.106.124036",
  journal = "Phys. Rev. D",
  volume = "106",
  number = "12",
  pages = "124036",
  year = "2022"}

@article{Ge:2024itl,
    author = "Ge, Bo-Xuan and Lim, Eugene A. and Sperhake, Ulrich and Evstafyeva, Tamara and Cors, Daniela and de Jong, Eloy and Croft, Robin and Helfer, Thomas",
    title = "{Hair is complicated: Gravitational waves from stable and unstable boson-star mergers}",
    eprint = "2410.23839",
    archivePrefix = "arXiv",
    primaryClass = "gr-qc",
    reportNumber = "KCL-TH-2024-59",
    month = "10",
    year = "2024",
    journal = ""
}

@article{Kaup:1968,
  title = {Klein-Gordon Geon},
  author = {Kaup, David J.},
  journal = {Phys. Rev.},
  volume = {172},
  issue = {5},
  pages = {1331--1342},
  numpages = {0},
  year = {1968},
  month = {Aug},
  publisher = {American Physical Society},
  doi = {10.1103/PhysRev.172.1331},
  url = {https://link.aps.org/doi/10.1103/PhysRev.172.1331}
}

@article{Ruffini:1969qy,
    author = "Ruffini, Remo and Bonazzola, Silvano",
    title = "{Systems of selfgravitating particles in general relativity and the concept of an equation of state}",
    doi = "10.1103/PhysRev.187.1767",
    journal = "Phys. Rev.",
    volume = "187",
    pages = "1767--1783",
    year = "1969"
}

@article{Sin_1994,
   title={Late-time phase transition and the galactic halo as a Bose liquid},
   volume={50},
   ISSN={0556-2821},
   url={http://dx.doi.org/10.1103/PhysRevD.50.3650},
   DOI={10.1103/physrevd.50.3650},
   number={6},
   journal={Physical Review D},
   publisher={American Physical Society (APS)},
   author={Sin, Sang-Jin},
   year={1994},
   month=sep, pages={3650–3654} }

@article{Schive_2014,
   title={Cosmic structure as the quantum interference of a coherent dark wave},
   volume={10},
   ISSN={1745-2481},
   url={http://dx.doi.org/10.1038/nphys2996},
   DOI={10.1038/nphys2996},
   number={7},
   journal={Nature Physics},
   publisher={Springer Science and Business Media LLC},
   author={Schive, Hsi-Yu and Chiueh, Tzihong and Broadhurst, Tom},
   year={2014},
   month=jun, pages={496–499} }

@article{Torres:2000dw,
    author = "Torres, Diego F. and Capozziello, S. and Lambiase, G.",
    title = "{A Supermassive scalar star at the galactic center?}",
    eprint = "astro-ph/0004064",
    archivePrefix = "arXiv",
    doi = "10.1103/PhysRevD.62.104012",
    journal = "Phys. Rev. D",
    volume = "62",
    pages = "104012",
    year = "2000"
}

@article{Guzman:2005bs,
    author = "Guzman, F. Siddhartha",
    title = "{Accretion disc onto boson stars: A Way to supplant black holes candidates}",
    eprint = "gr-qc/0512081",
    archivePrefix = "arXiv",
    doi = "10.1103/PhysRevD.73.021501",
    journal = "Phys. Rev. D",
    volume = "73",
    pages = "021501",
    year = "2006"
}

@article{Amaro_Seoane_2010,
   title={Constraining scalar fields with stellar kinematics and collisional dark matter},
   volume={2010},
   ISSN={1475-7516},
   url={http://dx.doi.org/10.1088/1475-7516/2010/11/002},
   DOI={10.1088/1475-7516/2010/11/002},
   number={11},
   journal={Journal of Cosmology and Astroparticle Physics},
   publisher={IOP Publishing},
   author={Amaro-Seoane, Pau and Barranco, Juan and Bernal, Argelia and Rezzolla, Luciano},
   year={2010},
   month=nov, pages={002–002} }

@article{Bezares_2022,
   title={Gravitational waves and kicks from the merger of unequal mass, highly compact boson stars},
   volume={105},
   ISSN={2470-0029},
   url={http://dx.doi.org/10.1103/PhysRevD.105.064067},
   DOI={10.1103/physrevd.105.064067},
   number={6},
   journal={Physical Review D},
   publisher={American Physical Society (APS)},
   author={Bezares, Miguel and Bošković, Mateja and Liebling, Steven and Palenzuela, Carlos and Pani, Paolo and Barausse, Enrico},
   year={2022},
   month=mar }

@article{Helfer_2022,
   title={Malaise and remedy of binary boson-star initial data},
   volume={39},
   ISSN={1361-6382},
   url={http://dx.doi.org/10.1088/1361-6382/ac53b7},
   DOI={10.1088/1361-6382/ac53b7},
   number={7},
   journal={Classical and Quantum Gravity},
   publisher={IOP Publishing},
   author={Helfer, Thomas and Sperhake, Ulrich and Croft, Robin and Radia, Miren and Ge, Bo-Xuan and Lim, Eugene A},
   year={2022},
   month=mar, pages={074001} }

@article{Lee:1987,
  title = {Soliton stars and the critical masses of black holes},
  author = {Lee, T. D.},
  journal = {Phys. Rev. D},
  volume = {35},
  issue = {12},
  pages = {3637--3639},
  numpages = {0},
  year = {1987},
  month = {Jun},
  publisher = {American Physical Society},
  doi = {10.1103/PhysRevD.35.3637},
  url = {https://link.aps.org/doi/10.1103/PhysRevD.35.3637}
}

@article{Guzman_2004, title={Evolving spherical boson stars on a 3D cartesian grid}, volume={70}, journal={Physical Review D}, author={Guzman, F. Siddhartha}, year={2004}, month=aug, pages={044033}, language={en} }

@article{Guzman_2009,
    author = "Guzm{\'a}n, F. S.",
    title = "{The three dynamical fates of Boson Stars}",
    eprint = "1907.08193",
    archivePrefix = "arXiv",
    primaryClass = "gr-qc",
    journal = "Rev. Mex. Fis.",
    volume = "55",
    pages = "321--326",
    year = "2009"
}

@article{Seidel_Suen_1990, title={Dynamical evolution of boson stars: Perturbing the ground state}, volume={42}, rights={http://link.aps.org/licenses/aps-default-license}, ISSN={0556-2821}, DOI={10.1103/PhysRevD.42.384}, number={2}, journal={Physical Review D}, author={Seidel, Edward and Suen, Wai-Mo}, year={1990}, month=jul, pages={384–403}, language={en} }

@misc{Marks:2025,
      title={Long-term stable nonlinear evolutions of ultracompact black-hole mimickers}, 
      author={Gareth Arturo Marks and Seppe J. Staelens and Tamara Evstafyeva and Ulrich Sperhake},
      year={2025},
      eprint={2504.17775},
      archivePrefix={arXiv},
      primaryClass={gr-qc},
      url={https://arxiv.org/abs/2504.17775}, 
}

@article{Franzin_2024,
   title={Mini boson stars in higher dimensions are radially unstable},
   volume={56},
   ISSN={1572-9532},
   url={http://dx.doi.org/10.1007/s10714-024-03287-9},
   DOI={10.1007/s10714-024-03287-9},
   number={9},
   journal={General Relativity and Gravitation},
   publisher={Springer Science and Business Media LLC},
   author={Franzin, Edgardo},
   year={2024},
   month=sep }

@article{Alcubierre_2019,
   title={Dynamical evolutions of $ \newcommand{\e}{{\rm e}
} \ell$ -boson stars in spherical symmetry},
   volume={36},
   ISSN={1361-6382},
   url={http://dx.doi.org/10.1088/1361-6382/ab4726},
   DOI={10.1088/1361-6382/ab4726},
   number={21},
   journal={Classical and Quantum Gravity},
   publisher={IOP Publishing},
   author={Alcubierre, Miguel and Barranco, Juan and Bernal, Argelia and Degollado, Juan Carlos and Diez-Tejedor, Alberto and Megevand, Miguel and Núñez, Darío and Sarbach, Olivier},
   year={2019},
   month=oct, pages={215013} }

@article{Sperhake_2001n,
  title = {Nonlinear radial oscillations of neutron stars},
  author = {Gabler, Michael and Sperhake, Ulrich and Andersson, Nils},
  journal = {Phys. Rev. D},
  volume = {80},
  issue = {6},
  pages = {064012},
  numpages = {16},
  year = {2009},
  month = {Sep},
  publisher = {American Physical Society},
  doi = {10.1103/PhysRevD.80.064012},
  url = {https://link.aps.org/doi/10.1103/PhysRevD.80.064012}
}

@article{Siemonsen_2021,
   title={Stability of rotating scalar boson stars with nonlinear interactions},
   volume={103},
   ISSN={2470-0029},
   url={http://dx.doi.org/10.1103/PhysRevD.103.044022},
   DOI={10.1103/physrevd.103.044022},
   number={4},
   journal={Physical Review D},
   publisher={American Physical Society (APS)},
   author={Siemonsen, Nils and East, William E.},
   year={2021},
   month=feb }

@article{Balakrishna_1998,
   title={Dynamical evolution of boson stars. II. Excited states and self-interacting fields},
   volume={58},
   ISSN={1089-4918},
   url={http://dx.doi.org/10.1103/PhysRevD.58.104004},
   DOI={10.1103/physrevd.58.104004},
   number={10},
   journal={Physical Review D},
   publisher={American Physical Society (APS)},
   author={Balakrishna, Jayashree and Seidel, Edward and Suen, Wai-Mo},
   year={1998},
   month=sep }

@article{Liebling_2023,
   title={Dynamical boson stars},
   volume={26},
   ISSN={1433-8351},
   url={http://dx.doi.org/10.1007/s41114-023-00043-4},
   DOI={10.1007/s41114-023-00043-4},
   number={1},
   journal={Living Reviews in Relativity},
   publisher={Springer Science and Business Media LLC},
   author={Liebling, Steven L. and Palenzuela, Carlos},
   year={2023},
   month=feb }

@article{Cardoso_2022,
   title={Piercing of a boson star by a black hole},
   volume={106},
   ISSN={2470-0029},
   url={http://dx.doi.org/10.1103/PhysRevD.106.044030},
   DOI={10.1103/physrevd.106.044030},
   number={4},
   journal={Physical Review D},
   publisher={American Physical Society (APS)},
   author={Cardoso, Vitor and Ikeda, Taishi and Zhong, Zhen and Zilhão, Miguel},
   year={2022},
   month=aug }

@article{Zhong_2023,
   title={Piercing of a solitonic boson star by a black hole},
   volume={108},
   ISSN={2470-0029},
   url={http://dx.doi.org/10.1103/PhysRevD.108.084051},
   DOI={10.1103/physrevd.108.084051},
   number={8},
   journal={Physical Review D},
   publisher={American Physical Society (APS)},
   author={Zhong, Zhen and Cardoso, Vitor and Ikeda, Taishi and Zilhão, Miguel},
   year={2023},
   month=oct }

@article{Olivares_2020, title={How to tell an accreting boson star from a black hole}, volume={497}, ISSN={0035-8711, 1365-2966}, DOI={10.1093/mnras/staa1878}, note={arXiv:1809.08682 [gr-qc]}, number={1}, journal={Monthly Notices of the Royal Astronomical Society}, author={Olivares, Hector and Younsi, Ziri and Fromm, Christian M. and Laurentis, Mariafelicia De and Porth, Oliver and Mizuno, Yosuke and Falcke, Heino and Kramer, Michael and Rezzolla, Luciano}, year={2020}, month=sep, pages={521–535}, language={en} }

@article{Lee_1989,
    title = {Stability of mini-boson stars},
    journal = {Nuclear Physics B},
    volume = {315},
    number = {2},
    pages = {477-516},
    year = {1989},
    issn = {0550-3213},
    doi = {https://doi.org/10.1016/0550-3213(89)90365-9},
    url = {https://www.sciencedirect.com/science/article/pii/0550321389903659},
    author = {T.D. Lee and Yang Pang}
    }

@article{Evstafyeva:2025mvx,
    author = "Evstafyeva, Tamara and Siemonsen, Nils and East, William E.",
    title = "{Assessing the stability of ultracompact spinning boson stars with nonlinear evolutions}",
    eprint = "2508.11527",
    archivePrefix = "arXiv",
    primaryClass = "gr-qc",
    month = "8",
    year = "2025"
}

@misc{Lai_2004,
      title={A Numerical Study of Boson Stars}, 
      author={Chi-Wai Lai},
      year={2004},
      eprint={gr-qc/0410040},
      archivePrefix={arXiv},
      primaryClass={gr-qc},
      url={https://arxiv.org/abs/gr-qc/0410040}, 
}

@misc{Lai_2007,
      title={Final Fate of Subcritical Evolutions of Boson Stars}, 
      author={Chi Wai Lai and Matthew W. Choptuik},
      year={2007},
      eprint={0709.0324},
      archivePrefix={arXiv},
      primaryClass={gr-qc},
      url={https://arxiv.org/abs/0709.0324}, 
}

@article{Choptuik_2010,
   title={Ultrarelativistic Particle Collisions},
   volume={104},
   ISSN={1079-7114},
   url={http://dx.doi.org/10.1103/PhysRevLett.104.111101},
   DOI={10.1103/physrevlett.104.111101},
   number={11},
   journal={Physical Review Letters},
   publisher={American Physical Society (APS)},
   author={Choptuik, Matthew W. and Pretorius, Frans},
   year={2010},
   month=mar }

@article{Bizo_2005,
   title={Critical Behavior in Vacuum Gravitational Collapse in<mml:math xmlns:mml="http://www.w3.org/1998/Math/MathML" display="inline"><mml:mn>4</mml:mn><mml:mo>+</mml:mo><mml:mn>1</mml:mn></mml:math>Dimensions},
   volume={95},
   ISSN={1079-7114},
   url={http://dx.doi.org/10.1103/PhysRevLett.95.071102},
   DOI={10.1103/physrevlett.95.071102},
   number={7},
   journal={Physical Review Letters},
   publisher={American Physical Society (APS)},
   author={Bizoń, Piotr and Chmaj, Tadeusz and Schmidt, Bernd G.},
   year={2005},
   month=aug }

@article{Porto_Veronese_2022,
   title={Critical phenomena in a gravitational collapse with a competing scalar field and gravitational waves in 
<mml:math xmlns:mml="http://www.w3.org/1998/Math/MathML" display="inline"><mml:mrow><mml:mn>4</mml:mn><mml:mo>+</mml:mo><mml:mn>1</mml:mn></mml:mrow></mml:math>
 dimensions},
   volume={106},
   ISSN={2470-0029},
   url={http://dx.doi.org/10.1103/PhysRevD.106.104044},
   DOI={10.1103/physrevd.106.104044},
   number={10},
   journal={Physical Review D},
   publisher={American Physical Society (APS)},
   author={Porto Veronese, Bernardo and Gundlach, Carsten},
   year={2022},
   month=nov }

@article{Yoo:2005nj,
    author = "Yoo, Chul-Moon and Nakao, Ken-ichi and Ida, Daisuke",
    title = "{Hoop conjecture in five-dimensions: Violation of cosmic censorship}",
    eprint = "gr-qc/0503008",
    archivePrefix = "arXiv",
    reportNumber = "OCU-PHYS-227, AP-GR-23",
    doi = "10.1103/PhysRevD.71.104014",
    journal = "Phys. Rev. D",
    volume = "71",
    pages = "104014",
    year = "2005"
}

@article{Blazquez-Salcedo:2019qrz,
    author = "Bl{\'a}zquez-Salcedo, Jose Luis and Knoll, Christian and Radu, Eugen",
    title = "{Boson and Dirac stars in $D\geq 4$ dimensions}",
    eprint = "1902.05851",
    archivePrefix = "arXiv",
    primaryClass = "gr-qc",
    doi = "10.1016/j.physletb.2019.04.035",
    journal = "Phys. Lett. B",
    volume = "793",
    pages = "161--168",
    year = "2019"
}

@article{Brihaye_2016,
   title={Minimal boson stars in 5 dimensions: classical instability and existence of ergoregions},
   volume={33},
   ISSN={1361-6382},
   url={http://dx.doi.org/10.1088/0264-9381/33/6/065002},
   DOI={10.1088/0264-9381/33/6/065002},
   number={6},
   journal={Classical and Quantum Gravity},
   publisher={IOP Publishing},
   author={Brihaye, Yves and Hartmann, Betti},
   year={2016},
   month=feb, pages={065002} }

@misc{Marks_2025_CP,
      title={Perturbations of Solitonic Boson Stars: Nonlinear Radial Stability and Binding Energy}, 
      author={Gareth Arturo Marks},
      year={2025},
      eprint={2508.11757},
      archivePrefix={arXiv},
      primaryClass={gr-qc},
      url={https://arxiv.org/abs/2508.11757}, 
}

@article{Di_Giovanni_2020,
   title={Dynamical bar-mode instability in spinning bosonic stars},
   volume={102},
   ISSN={2470-0029},
   url={http://dx.doi.org/10.1103/PhysRevD.102.124009},
   DOI={10.1103/physrevd.102.124009},
   number={12},
   journal={Physical Review D},
   publisher={American Physical Society (APS)},
   author={Di Giovanni, Fabrizio and Sanchis-Gual, Nicolas and Cerdá-Durán, Pablo and Zilhão, Miguel and Herdeiro, Carlos and Font, José A. and Radu, Eugen},
   year={2020},
   month=dec }

@misc{Brito_2025,
      title={Stability and collisions of excited spherical boson stars: glimpses of chains and rings}, 
      author={Marco Brito and Carlos Herdeiro and Eugen Radu and Nicolas Sanchis-Gual and Miguel Zilhão},
      year={2025},
      eprint={2506.06442},
      archivePrefix={arXiv},
      primaryClass={gr-qc},
      url={https://arxiv.org/abs/2506.06442}, 
}

@article{Pretorius_2004, title={Numerical Relativity Using a Generalized Harmonic Decomposition}, volume={22}, ISSN={0264-9381, 1361-6382}, DOI={10.1088/0264-9381/22/2/014}, note={arXiv:gr-qc/0407110}, number={2}, journal={Classical and Quantum Gravity}, author={Pretorius, Frans}, year={2005}, month=jan, pages={425–451}, language={en} }

@article{Evstafyeva_2023,
   title={Unequal-mass boson-star binaries: initial data and merger dynamics},
   volume={40},
   ISSN={1361-6382},
   url={http://dx.doi.org/10.1088/1361-6382/acc2a8},
   DOI={10.1088/1361-6382/acc2a8},
   number={8},
   journal={Classical and Quantum Gravity},
   publisher={IOP Publishing},
   author={Evstafyeva, Tamara and Sperhake, Ulrich and Helfer, Thomas and Croft, Robin and Radia, Miren and Ge, Bo-Xuan and Lim, Eugene A},
   year={2023},
   month=mar, pages={085009} }

@article{Gleiser:1988rq,
    author = "Gleiser, Marcelo",
    title = "{Stability of Boson Stars}",
    reportNumber = "FERMILAB-PUB-88-067-A",
    doi = "10.1103/PhysRevD.38.2376",
    journal = "Phys. Rev. D",
    volume = "38",
    pages = "2376",
    year = "1988",
    note = "[Erratum: Phys.Rev.D 39, 1257 (1989)]"
}

@article{Hawley_2000,
   title={Boson stars driven to the brink of black hole formation},
   volume={62},
   ISSN={1089-4918},
   url={http://dx.doi.org/10.1103/PhysRevD.62.104024},
   DOI={10.1103/physrevd.62.104024},
   number={10},
   journal={Physical Review D},
   publisher={American Physical Society (APS)},
   author={Hawley, Scott H. and Choptuik, Mathew W.},
   year={2000},
   month=oct }

@article{Santos:2024vdm,
    author = "Santos, Nuno M. and Benone, Carolina L. and Herdeiro, Carlos A. R.",
    title = "{Radial stability of spherical bosonic stars and critical points}",
    eprint = "2404.07257",
    archivePrefix = "arXiv",
    primaryClass = "gr-qc",
    doi = "10.1088/1475-7516/2024/06/068",
    journal = "JCAP",
    volume = "06",
    pages = "068",
    year = "2024"
}

@article{Kusmartsev:1990cr,
    author = "Kusmartsev, Fjodor V. and Mielke, Eckehard W. and Schunck, Franz E.",
    title = "{Gravitational stability of boson stars}",
    eprint = "0810.0696",
    archivePrefix = "arXiv",
    primaryClass = "astro-ph",
    reportNumber = "PRINT-90-0362 (COLOGNE)",
    doi = "10.1103/PhysRevD.43.3895",
    journal = "Phys. Rev. D",
    volume = "43",
    pages = "3895--3901",
    year = "1991"
}

@article{Sanchis_Gual_2022,
   title={Self-interactions can stabilize excited boson stars},
   volume={39},
   ISSN={1361-6382},
   url={http://dx.doi.org/10.1088/1361-6382/ac4b9b},
   DOI={10.1088/1361-6382/ac4b9b},
   number={6},
   journal={Classical and Quantum Gravity},
   publisher={IOP Publishing},
   author={Sanchis-Gual, Nicolas and Herdeiro, Carlos and Radu, Eugen},
   year={2022},
   month=feb, pages={064001} }

@article{Bona_1995,
   title={New Formalism for Numerical Relativity},
   volume={75},
   ISSN={1079-7114},
   url={http://dx.doi.org/10.1103/PhysRevLett.75.600},
   DOI={10.1103/physrevlett.75.600},
   number={4},
   journal={Physical Review Letters},
   publisher={American Physical Society (APS)},
   author={Bona, Carles and Massó, Joan and Seidel, Edward and Stela, Joan},
   year={1995},
   month=jul, pages={600–603} }

@article{Alcubierre_1997,
   title={Appearance of coordinate shocks in hyperbolic formalisms of general relativity},
   volume={55},
   ISSN={1089-4918},
   url={http://dx.doi.org/10.1103/PhysRevD.55.5981},
   DOI={10.1103/physrevd.55.5981},
   number={10},
   journal={Physical Review D},
   publisher={American Physical Society (APS)},
   author={Alcubierre, Miguel},
   year={1997},
   month=may, pages={5981–5991} }

@article{Astefanesei_2003,
   title={Boson stars with negative cosmological constant},
   volume={665},
   ISSN={0550-3213},
   url={http://dx.doi.org/10.1016/S0550-3213(03)00482-6},
   DOI={10.1016/s0550-3213(03)00482-6},
   journal={Nuclear Physics B},
   publisher={Elsevier BV},
   author={Astefanesei, Dumitru and Radu, Eugen},
   year={2003},
   month=aug, pages={594–622} }

@article{Sennett_2017,
   title={Distinguishing boson stars from black holes and neutron stars from tidal interactions in inspiraling binary systems},
   volume={96},
   ISSN={2470-0029},
   url={http://dx.doi.org/10.1103/PhysRevD.96.024002},
   DOI={10.1103/physrevd.96.024002},
   number={2},
   journal={Physical Review D},
   publisher={American Physical Society (APS)},
   author={Sennett, Noah and Hinderer, Tanja and Steinhoff, Jan and Buonanno, Alessandra and Ossokine, Serguei},
   year={2017},
   month=jul }

@article{Liang_2025,
   title={Bifurcations in bosonic stars: chains and rings from spherical solutions},
   volume={2025},
   ISSN={1029-8479},
   url={http://dx.doi.org/10.1007/JHEP03(2025)119},
   DOI={10.1007/jhep03(2025)119},
   number={3},
   journal={Journal of High Energy Physics},
   publisher={Springer Science and Business Media LLC},
   author={Liang, Chen and Herdeiro, Carlos A. R. and Radu, Eugen},
   year={2025},
   month=mar }

@article{Brito_2016,
   title={Proca stars: Gravitating Bose–Einstein condensates of massive spin 1 particles},
   volume={752},
   ISSN={0370-2693},
   url={http://dx.doi.org/10.1016/j.physletb.2015.11.051},
   DOI={10.1016/j.physletb.2015.11.051},
   journal={Physics Letters B},
   publisher={Elsevier BV},
   author={Brito, Richard and Cardoso, Vitor and Herdeiro, Carlos A.R. and Radu, Eugen},
   year={2016},
   month=jan, pages={291–295} }

@article{Palenzuela_2017,
   title={Gravitational wave signatures of highly compact boson star binaries},
   volume={96},
   ISSN={2470-0029},
   url={http://dx.doi.org/10.1103/PhysRevD.96.104058},
   DOI={10.1103/physrevd.96.104058},
   number={10},
   journal={Physical Review D},
   publisher={American Physical Society (APS)},
   author={Palenzuela, Carlos and Pani, Paolo and Bezares, Miguel and Cardoso, Vitor and Lehner, Luis and Liebling, Steven},
   year={2017},
   month=nov }

@article{Emparan_2013,
   title={The large D limit of General Relativity},
   volume={2013},
   ISSN={1029-8479},
   url={http://dx.doi.org/10.1007/JHEP06(2013)009},
   DOI={10.1007/jhep06(2013)009},
   number={6},
   journal={Journal of High Energy Physics},
   publisher={Springer Science and Business Media LLC},
   author={Emparan, Roberto and Suzuki, Ryotaku and Tanabe, Kentaro},
   year={2013},
   month=jun }

@article{Cook_2017,
   title={Black-hole head-on collisions in higher dimensions},
   volume={96},
   ISSN={2470-0029},
   url={http://dx.doi.org/10.1103/PhysRevD.96.124006},
   DOI={10.1103/physrevd.96.124006},
   number={12},
   journal={Physical Review D},
   publisher={American Physical Society (APS)},
   author={Cook, William G. and Sperhake, Ulrich and Berti, Emanuele and Cardoso, Vitor},
   year={2017},
   month=dec }

@article{Bustillo_2021,
   title={GW190521 as a Merger of Proca Stars: A Potential New Vector Boson of 
<mml:math xmlns:mml="http://www.w3.org/1998/Math/MathML" display="inline"><mml:mrow><mml:mn>8.7</mml:mn><mml:mo>×</mml:mo><mml:msup><mml:mrow><mml:mn>10</mml:mn></mml:mrow><mml:mrow><mml:mo>−</mml:mo><mml:mn>13</mml:mn></mml:mrow></mml:msup><mml:mtext>  </mml:mtext><mml:mi>eV</mml:mi></mml:mrow></mml:math>},
   volume={126},
   ISSN={1079-7114},
   url={http://dx.doi.org/10.1103/PhysRevLett.126.081101},
   DOI={10.1103/physrevlett.126.081101},
   number={8},
   journal={Physical Review Letters},
   publisher={American Physical Society (APS)},
   author={Bustillo, Juan Calderón and Sanchis-Gual, Nicolas and Torres-Forné, Alejandro and Font, José A. and Vajpeyi, Avi and Smith, Rory and Herdeiro, Carlos and Radu, Eugen and Leong, Samson H. W.},
   year={2021},
   month=feb }

@article{Sanchis_Gual_2022_proca,
   title={Impact of the wavelike nature of Proca stars on their gravitational-wave emission},
   volume={106},
   ISSN={2470-0029},
   url={http://dx.doi.org/10.1103/PhysRevD.106.124011},
   DOI={10.1103/physrevd.106.124011},
   number={12},
   journal={Physical Review D},
   publisher={American Physical Society (APS)},
   author={Sanchis-Gual, Nicolas and Bustillo, Juan Calderón and Herdeiro, Carlos and Radu, Eugen and Font, José A. and Leong, Samson H. W. and Torres-Forné, Alejandro},
   year={2022},
   month=dec }

@article{Cardoso_2022_eco,
   title={ECO-spotting: looking for extremely compact objects with bosonic fields},
   volume={39},
   ISSN={1361-6382},
   url={http://dx.doi.org/10.1088/1361-6382/ac41e7},
   DOI={10.1088/1361-6382/ac41e7},
   number={3},
   journal={Classical and Quantum Gravity},
   publisher={IOP Publishing},
   author={Cardoso, Vitor and Macedo, Caio F B and Maeda, Kei-ichi and Okawa, Hirotada},
   year={2022},
   month=jan, pages={034001} }

@misc{Marks_Spherical_BS_Evolver,
  author       = {Gareth A. Marks},
  title        = {Spherical\_BS\_Evolver},
  year         = {2025},
  publisher    = {GitHub},
  journal      = {GitHub repository},
  howpublished = {\url{https://github.com/GarethAMarks/Spherical_BS_Evolver/tree/master}},
  note         = {Accessed: 2025-10-09}
}

@article{Rosa_2022_mim,
   title={Shadows of boson and Proca stars with thin accretion disks},
   volume={106},
   ISSN={2470-0029},
   url={http://dx.doi.org/10.1103/PhysRevD.106.084004},
   DOI={10.1103/physrevd.106.084004},
   number={8},
   journal={Physical Review D},
   publisher={American Physical Society (APS)},
   author={Rosa, João Luís and Rubiera-Garcia, Diego},
   year={2022},
   month=oct }

@article{Rosa_2023,
   title={Imaging compact boson stars with hot spots and thin accretion disks},
   volume={108},
   ISSN={2470-0029},
   url={http://dx.doi.org/10.1103/PhysRevD.108.044021},
   DOI={10.1103/physrevd.108.044021},
   number={4},
   journal={Physical Review D},
   publisher={American Physical Society (APS)},
   author={Rosa, João Luís and Macedo, Caio F. B. and Rubiera-Garcia, Diego},
   year={2023},
   month=aug }

@misc{Rosa_2025,
      title={Polarimetry imprints of exotic compact objects: solitonic boson stars}, 
      author={João Luís Rosa and Nicolas Aimar and Hanna Liis Tamm},
      year={2025},
      eprint={2504.02472},
      archivePrefix={arXiv},
      primaryClass={gr-qc},
      url={https://arxiv.org/abs/2504.02472}, 
}

@article{Siemonsen_2023,
   title={Generic initial data for binary boson stars},
   volume={108},
   ISSN={2470-0029},
   url={http://dx.doi.org/10.1103/PhysRevD.108.124015},
   DOI={10.1103/physrevd.108.124015},
   number={12},
   journal={Physical Review D},
   publisher={American Physical Society (APS)},
   author={Siemonsen, Nils and East, William E.},
   year={2023},
   month=dec }

@inproceedings{Staelens:2025wom,
    author = "Staelens, Seppe J.",
    title = "{Search for growing angular modes in ultracompact boson star evolutions}",
    eprint = "2508.19460",
    archivePrefix = "arXiv",
    primaryClass = "gr-qc",
    month = "8",
    year = "2025"
}

@article{Mourelle_24,
  title = {Galactic halos and rotating bosonic dark matter},
  author = {Mourelle, Jorge Castelo and Adam, Christoph},
  journal = {Phys. Rev. D},
  volume = {110},
  issue = {12},
  pages = {123001},
  numpages = {19},
  year = {2024},
  month = {Dec},
  publisher = {American Physical Society},
  doi = {10.1103/PhysRevD.110.123001},
  url = {https://link.aps.org/doi/10.1103/PhysRevD.110.123001}
}

@article{Mourelle_25,
  title = {Reproducing galactic rotation curves with a two-component bosonic dark matter model},
  author = {Mourelle, Jorge Castelo and Sanchis-Gual, Nicolas and Font, Jos\'e A.},
  journal = {Phys. Rev. D},
  volume = {112},
  issue = {6},
  pages = {063054},
  numpages = {16},
  year = {2025},
  month = {Sep},
  publisher = {American Physical Society},
  doi = {10.1103/13hw-msfg},
  url = {https://link.aps.org/doi/10.1103/13hw-msfg}
}

@article{Antoniadis:1990ew,
    author = "Antoniadis, Ignatios",
    title = "{A Possible new dimension at a few TeV}",
    reportNumber = "EP-CPTH-A978-0690",
    doi = "10.1016/0370-2693(90)90617-F",
    journal = "Phys. Lett. B",
    volume = "246",
    pages = "377--384",
    year = "1990"
}

@article{Antoniadis_1998,
   title={New dimensions at a millimeter to a fermi and superstrings at a TeV},
   volume={436},
   ISSN={0370-2693},
   url={http://dx.doi.org/10.1016/S0370-2693(98)00860-0},
   DOI={10.1016/s0370-2693(98)00860-0},
   number={3–4},
   journal={Physics Letters B},
   publisher={Elsevier BV},
   author={Antoniadis, Ignatios and Arkani-Hamed, Nima and Dimopoulos, Savas and Dvali, Gia},
   year={1998},
   month=sep, pages={257–263} }

@article{Arkani_Hamed_1998,
   title={The hierarchy problem and new dimensions at a millimeter},
   volume={429},
   ISSN={0370-2693},
   url={http://dx.doi.org/10.1016/S0370-2693(98)00466-3},
   DOI={10.1016/s0370-2693(98)00466-3},
   number={3–4},
   journal={Physics Letters B},
   publisher={Elsevier BV},
   author={Arkani–Hamed, Nima and Dimopoulos, Savas and Dvali, Gia},
   year={1998},
   month=jun, pages={263–272} }

@article{Witek_2011,
  title = {Head-on collisions of unequal mass black holes in $D=5$ dimensions},
  author = {Witek, Helvi and Cardoso, Vitor and Gualtieri, Leonardo and Herdeiro, Carlos and Sperhake, Ulrich and Zilh\~ao, Miguel},
  journal = {Phys. Rev. D},
  volume = {83},
  issue = {4},
  pages = {044017},
  numpages = {6},
  year = {2011},
  month = {Feb},
  publisher = {American Physical Society},
  doi = {10.1103/PhysRevD.83.044017},
  url = {https://link.aps.org/doi/10.1103/PhysRevD.83.044017}
}

@article{Sperhake_2019,
   title={High-energy collision of black holes in higher dimensions},
   volume={100},
   ISSN={2470-0029},
   url={http://dx.doi.org/10.1103/PhysRevD.100.104046},
   DOI={10.1103/physrevd.100.104046},
   number={10},
   journal={Physical Review D},
   publisher={American Physical Society (APS)},
   author={Sperhake, Ulrich and Cook, William and Wang, Diandian},
   year={2019},
   month=nov }

@article{Emparan_2008,
   title={Black Holes in Higher Dimensions},
   volume={11},
   ISSN={1433-8351},
   url={http://dx.doi.org/10.12942/lrr-2008-6},
   DOI={10.12942/lrr-2008-6},
   number={1},
   journal={Living Reviews in Relativity},
   publisher={Springer Science and Business Media LLC},
   author={Emparan, Roberto and Reall, Harvey S.},
   year={2008},
   month=sep }

\end{document}